\documentclass[
  reprint,
  amsmath,amssymb,
  aps,
  prmaterials,            
  superscriptaddress,
  longbibliography,
  floatfix
]{revtex4-2}

\usepackage{chemformula}
\usepackage{graphicx}
\usepackage{dcolumn}
\usepackage{bm}
\usepackage{xcolor}
\usepackage[T1]{fontenc}
\usepackage[utf8]{inputenc}
\usepackage{textcomp}
\usepackage{natbib}

\usepackage{siunitx}

\usepackage{booktabs}

\usepackage{array}
\usepackage{multirow}
\usepackage{longtable}
\usepackage{pdfpages}
\usepackage{hyperref}
\usepackage{atbegshi}
\usepackage{csvsimple}
\usepackage{seqsplit}
\usepackage{etoolbox}
\usepackage{pdflscape}
\usepackage{xparse}
\newcolumntype{P}[1]{>{\raggedright\arraybackslash}p{#1}}

\newcommand{\SymprecFmt}[1]{%
  \num[scientific-notation=true,print-unity-mantissa=false]{#1}%
}

\AtBeginShipout{\pdfpageattr{/Rotate 0}}
\includepdfset{pagecommand=\thispagestyle{empty},turn=false}
\AtBeginDocument{%
  \global\pdfpagewidth=\paperwidth
  \global\pdfpageheight=\paperheight
}

\newcounter{S2row}
\newcounter{S3row}

\makeatletter
\AtBeginDocument{\let\LS@rot\@undefined}
\makeatother

\newcommand{\NumOrDashPlain}[1]{%
  \edef\temp{#1}%
  \ifblank{#1}{--}{%
    \ifdefstring{\temp}{nan}{--}{%
    \ifdefstring{\temp}{NaN}{--}{%
    \ifdefstring{\temp}{inf}{--}{%
    \ifdefstring{\temp}{Inf}{--}{%
    \ifdefstring{\temp}{-inf}{--}{%
    \ifdefstring{\temp}{-Inf}{--}{%
      \num[
        scientific-notation=false,
        round-mode=places,
        round-precision=3
      ]{#1}%
    }}}}}}%
  }%
}

\newcommand{\NumOrDashSci}[1]{%
  \edef\temp{#1}%
  \ifblank{#1}{--}{%
    \ifdefstring{\temp}{nan}{--}{%
    \ifdefstring{\temp}{NaN}{--}{%
    \ifdefstring{\temp}{inf}{--}{%
    \ifdefstring{\temp}{Inf}{--}{%
    \ifdefstring{\temp}{-inf}{--}{%
    \ifdefstring{\temp}{-Inf}{--}{%
      \num[
        scientific-notation=true,
        round-mode=places,
        round-precision=3
      ]{#1}%
    }}}}}}%
  }%
}

\ExplSyntaxOn
\NewDocumentCommand{\SGfmt}{m}
 {
  \tl_set:Nx \l_tmpa_tl { \tl_to_str:e { #1 } }
  \regex_replace_all:nnN { ["“”] } { } \l_tmpa_tl
  \regex_replace_all:nnN { _\s*([0-9]+) } { $_{\1}$ } \l_tmpa_tl
  \tl_use:N \l_tmpa_tl
 }
\ExplSyntaxOff

\begin{document}
\title{High-Throughput Quantification of Altermagnetic Band Splitting}

\author{Ali Sufyan}
\affiliation{NanoLund and Division of Mathematical Physics, Department of Physics, Lund University, SE-221 00 Lund, Sweden}
\affiliation{Wallenberg Initiative Materials Science for Sustainability, Department of Physics, Lund University, SE-221 00 Lund, Sweden}
\affiliation{Applied Physics, Division of Materials Science, Department of Engineering Sciences and Mathematics, Lule\aa\ University of Technology, SE-971 87 Lule\aa\, Sweden}
\author{Brahim Marfoua}
\affiliation{Department of Physics, Chemistry and Biology, Link\"oping University, SE-581 83 Link\"oping, Sweden}
\author{J. Andreas Larsson}
\affiliation{Applied Physics, Division of Materials Science, Department of Engineering Sciences and Mathematics, Lule\aa\ University of Technology, SE-971 87 Lule\aa\, Sweden}
\affiliation{Wallenberg Initiative Materials Science for Sustainability, Lule\aa\ University of Technology, SE-971 87 Lule\aa\, Sweden}
\author{Erik van Loon}
\affiliation{NanoLund and Division of Mathematical Physics, Department of Physics, Lund University, SE-221 00 Lund, Sweden}
\affiliation{Wallenberg Initiative Materials Science for Sustainability, Department of Physics, Lund University, SE-221 00 Lund, Sweden}
\email{erik.van_loon@fysik.lu.se}
\author{Rickard Armiento}
\affiliation{Department of Physics, Chemistry and Biology, Link\"oping University, SE-581 83 Link\"oping, Sweden}
\email{rickard.armiento@liu.se}

\begin{abstract} 
Altermagnetism represents a recently established class of collinear magnetism that combines zero net magnetization with momentum-dependent spin polarization, enabled by symmetry constraints rather than spin-orbit coupling. This distinctive behavior gives rise to sizable spin splitting even in materials composed of light, earth-abundant elements, offering promising prospects for next-generation spintronics applications. Here, we present a comprehensive high-throughput screening of the 2287 entiries comprising the MAGNDATA database, integrating symmetry analysis with spin-polarized density functional theory (DFT) calculations to identify and characterize altermagnetic candidates. Our workflow investigates the collinear structures in the data set and collinear versions of the ones reported to be non-collinear, uncovering 180 materials exhibiting significant spin splitting, spanning both metallic and semiconducting systems. Detailed results are available for all 180 materials, but we also particularly discuss $\mathrm{UCr}_2\mathrm{Si}_2\mathrm{C}$, $\mathrm{NbMnP}$, and $\mathrm{YRuO}_3$ as representative cases with large spin splitting. Furthermore, comparison with the Computational 2D Materials Database (C2DB) and the AiiDA 2D repository gives 9 bulk altermagnets with chemically equivalent 2D counterparts linked to the same ICSD parent entry. Crucially, our momentum-resolved analysis reveals that the spin splitting varies strongly across the Brillouin zone, and that the maximal splitting tends to occur away from the high-symmetry paths, a result that directly informs and guides future photoemission experiments. 
By expanding the catalog of known altermagnets, this work lays a robust foundation for future experimental and theoretical advances in spintronics and quantum materials discovery.

\end{abstract}

\maketitle

\par 
\section{Introduction}
Altermagnetism has recently emerged as a distinct magnetic phase that expands beyond the traditional classifications of ferromagnetism (FM) and antiferromagnetism (AFM) \cite{vsmejkal2022beyond, vsmejkal2022emerging, mazin2022altermagnetism}. The concept was initially introduced through theoretical predictions in rutile AFM RuO$_2$ in 2019 \cite{ahn2019antiferromagnetism}, and subsequently gained significant attention through systematic investigations of its distinctive characteristics in 2022, at which point the term ``altermagnetism'' was formally established \cite{vsmejkal2022emerging, vsmejkal2022beyond}. Conventional FM materials exhibit a net magnetization due to parallel alignment of spins, breaking time-reversal symmetry (TRS) uniformly throughout the crystal. In contrast, traditional AFMs have globally compensated spin arrangements with adjacent spins oriented antiparallel, thereby generating no net magnetization and typically enforcing spin degeneracy in their electronic band structure due to symmetry operations combining inversion or translational symmetry with TRS. 

Altermagnets, however, possess globally compensated magnetization similar to AFMs but uniquely exhibit momentum-dependent spin polarization reminiscent of FM materials \cite{vsmejkal2022beyond, vsmejkal2022emerging, mazin2022altermagnetism, naka2019spin, vsmejkal2020crystal, ahn2019antiferromagnetism, yuan2020giant, hayami2020bottom, ma2021multifunctional, mazin2021prediction, liu2022spin, rooj2025altermagnetism}. The essential distinction arises from the symmetry relations between their spin sublattices: in altermagnets, opposite-spin sublattices are related by rotation symmetries combined with TRS, rather than translation or inversion symmetries. This crucial difference in symmetry leads to nonrelativistic spin splitting of electronic bands, breaking Kramers degeneracy without the necessity of spin-orbit coupling (SOC) \cite{yuan2020giant, vsmejkal2020crystal, ahn2019antiferromagnetism, hayami2019momentum, mazin2021prediction, krempasky2024altermagnetic}. In conventional AFMs, symmetry-enforced degeneracies typically prevent such splitting, while altermagnetic symmetry permits momentum-dependent spin-split bands that retain overall global spin compensation. Critically, because this spin splitting does not require SOC, altermagnetic behavior can manifest significantly even in materials composed of lighter, more earth-abundant elements like Fe and Mn. This broadens the scope for material discovery and significantly reduces reliance on heavier, expensive, or rare elements typically needed for SOC-driven phenomena \cite{wei2024crystal}. Experimentally confirmed altermagnets now encompass diverse systems, from insulating compounds like CuF$_2$ \cite{vsmejkal2022beyond}, MnF$_2$ \cite{yuan2020giant}, and MnTe \cite{krempasky2024altermagnetic} to metallic materials such as RuO$_2$ \cite{fedchenko2024observation}, Mn$_5$Si$_3$ \cite{reichlova2024observation}, and CrSb \cite{reimers2024direct}. These materials have demonstrated compelling and technologically relevant phenomena, including anomalous Hall effects (AHE), spin-polarized conductivity, spin-transfer torque, tunneling magnetoresistance, and giant magnetoresistance, which are essential ingredients for future spintronic applications \cite{vsmejkal2022emerging, vsmejkal2022anomalous, gonzalez2023spontaneous, vsmejkal2022giant, gonzalez2021efficient, bai2023efficient, bose2022tilted}.  Consequently, the identification and characterization of altermagnetic materials represent a transformative advancement in magnetism and spintronics, sparking intensive research efforts aimed at uncovering new materials, elucidating underlying physical principles, and developing practical applications for this emerging class of magnetic materials.  

However, discovering new altermagnets with significant spin splitting through trial-and-error experimental approaches is inherently challenging and resource-intensive. To overcome this limitation, high-throughput computational screening, leveraging reliable first-principles density functional theory (DFT) calculations, has emerged as an efficient and powerful methodology. This approach enables the rapid and systematic exploration of large material databases for candidates exhibiting desirable characteristics. Some recent efforts have used the MAGNDATA database~\cite{gallego2016magndata1, gallego2016magndata2} for this purpose (see Sec.~\ref{sec:comparison} for a comparison) with a focus on specific subclasses, such as metallic altermagnets or magnonic excitations in collinear systems, or using other selection criteria~\cite{wan2024high,chen2025unconventional,Guo2023}. In this present work, we undertake an exhaustive and systematic high-throughput screening of the entire MAGNDATA database, including collinear versions of the non-collinear entries, without constraints to specific classes or properties, aiming to identify and comprehensively characterize all potential altermagnetic candidates. Through this extensive computational investigation, we successfully discover numerous previously unreported altermagnetic materials, validate several known candidates, and provide robust theoretical frameworks for future experimental realizations. This comprehensive approach not only significantly enhances the current understanding of altermagnetic phenomena but also expands the available pool of materials for next-generation spintronic devices, highlighting the critical importance and novelty of this study.

\section{Screening workflow}
\begin{figure*}[t]
  \centering
  \includegraphics[width=\textwidth,trim=5pt 3pt 6pt 6pt,clip]{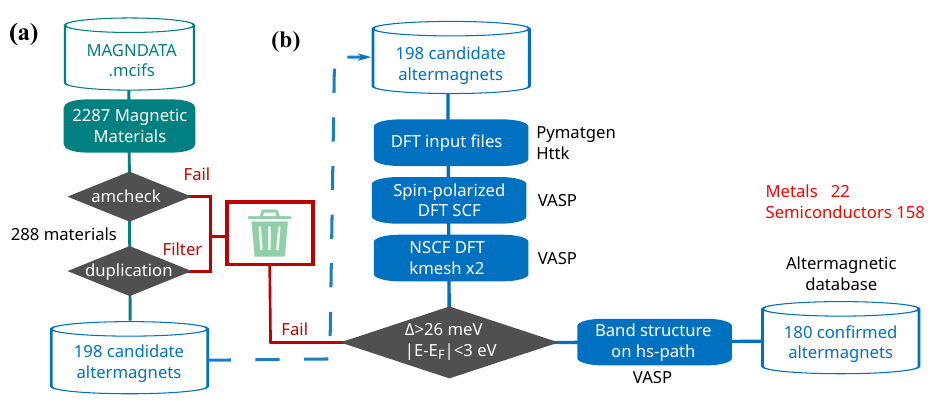}  
    \caption{\textbf{Two-stage high-throughput computational workflow to identify candidate altermagnetic materials (AMs).} \textbf{(a)} Initial screening selects experimentally characterized magnetic materials from the MAGNDATA database. Symmetry analysis via amcheck followed by deduplication filtering step, yielding a final set of 198 unique AM candidates. \textbf{(b)} In-depth computational verification using spin-polarized DFT calculations (VASP, pymatgen, and \textit{httk}). Materials exhibiting spin splitting ($\ge$ 26 meV within $\pm$3 eV of $E_F$) are confirmed as AM and included in the final database.}
    \label{fig:fig1}
\end{figure*}

We present a rigorous two-stage high-throughput computational screening workflow for the systematic identification of candidate altermagnetic materials, summarized schematically in Fig.~\ref{fig:fig1}. In the first stage of screening (Fig.~\ref{fig:fig1}a), we retrieved the March 11, 2025 state of the MAGNDATA database hosted by the Bilbao Crystallographic Server \cite{gallego2016magndata1, gallego2016magndata2, Aroyo2011}. This database contains 2287 published magnetic structures with crystal structures and magnetic configurations primarily obtained via experimental characterization.

These magnetic materials were subsequently analyzed using the recently developed computational tool amcheck \cite{smolyanyuk2024tool}. Amcheck operationalizes the symmetry-based theoretical framework underpinning altermagnetism by assessing symmetry operations relating spin-up and spin-down magnetic sublattices. Specifically, the algorithm evaluates whether the magnetic sublattices can be mapped onto each other through a spatial inversion or lattice translation combined with a spin-flip operation. If such mappings are universally applicable to all magnetic atoms, the material is pure AFM since these symmetry operations guarantee spin degeneracy across the electronic band structure. Conversely, the absence of such inversion or translation symmetries, coupled with the presence of alternative symmetry operations (e.g., rotational or mirror symmetries), indicates altermagnetism. This condition implies the presence of symmetry-protected spin splitting in the band structure despite zero net magnetization. The advantage of amcheck lies in its reliance solely on crystallographic and magnetic structural information, circumventing the need for computationally intensive electronic structure calculations. Consequently, it serves as a rapid, efficient, and theoretically robust tool ideally suited to large-scale, high-throughput screenings. The symmetry analysis, both internally in amcheck and for the details we report in the supplementary materials uses Spglib \cite{spglib, spglibv2}.

We used Pymatgen \cite{ong2013python} to read the mcif files from MAGNDATA. We applied amcheck to all the collinear structures, which resulted in 157 candidate altermagnetic materials. We then applied amcheck to collinear versions of all noncollinear materials, where we aligned the spins by the algorithm used by Pymatgen when converting a magnetic moment to a scalar: the magnitude of the magnetic moment is retained and the sign assigned by projecting the moment onto a standard axis. The standard axis is arbitrarily chosen to always have the direction defined by the Cartesian vector $(x,y,z) = (1.01, 1.02, 1.03)$. Notably, the forced-collinear variants are introduced solely to enable a uniform high-throughput screening; we do not assess their energetic stability relative to the noncollinear ground state.
%
This resulted in 131 additional candidates. 
Overall, we obtained a total of 288 potential AM materials from the MAGNDATA database (157 collinear and 131 noncollinear). 
Eliminating duplicated structures based on chemical composition and space group, the final set was 188 candidates suitable for further computational examination.

In the second screening phase (Fig.~\ref{fig:fig1}b), input files for spin-polarized self-consistent field (SCF) and non-self-consistent field (non-SCF) DFT calculations were systematically generated using Pymatgen and the High-throughput toolkit (\textit{httk}) \cite{httk}. The workflow manager in \textit{httk} provides a powerful, flexible automation framework specifically tailored for large-scale computational workflows, significantly enhancing computational efficiency. Importantly, magnetic orders and initial magnetic moments of atoms were directly extracted from MAGNDATA's 
mcif files, ensuring consistency with experimentally established configurations. In cases where the 
reported local magnetic moments were particularly small (e.g., below 1~$\mu_\mathrm{B}$/atom), we initialized the corresponding magnetic moments to a slightly higher value (typically 1~$\mu_\mathrm{B}$) to avoid artificial quenching during SCF relaxation. 

The resulting spin-polarized SCF calculations, performed using the Vienna Ab-initio Simulation Package (VASP)~\cite{paw1,paw2, vasp}, were analyzed to identify meaningful spin-splitting. To select only robust altermagnets, a minimum spin-splitting magnitude of 26 meV was used as a selection criterion and this splitting has to occur within an energy window of ±3 eV around the Fermi level. Figure S1 shows the distribution of $\Delta_{\mathrm{max}}$ across the screened set and supports adopting 26 meV as a physically motivated lower bound, comparable to the room-temperature thermal energy scale ($k_BT \approx 26$ meV).

Based on these criteria, 180 materials were validated as exhibiting significant spin-splitting and advanced to the final characterization step, while the remaining candidates were discarded. Detailed electronic band structure calculations were subsequently conducted for these selected 180 altermagnets, enabling direct visualization and confirmation of spin-up and spin-down band splittings. Because magnetic symmetry can enforce nodal manifolds where the nonrelativistic splitting vanishes on many high-symmetry lines/planes, we evaluate $\Delta(\mathbf{k})$ on dense Brillouin-zone meshes (and fixed-$k_z$ slices when needed), rather than relying solely on conventional high-symmetry paths, to reliably capture $\Delta_{\max}$. In these computations, the magnetic moments were assumed to be in a collinear configuration, a commonly adopted simplification in high-throughput studies, which is sufficient for identifying the essential altermagnetic characteristics. 

To further quantify the spin splitting, we computed both the average and volumetric spin-splitting metrics. The average spin splitting $\langle \Delta \rangle$ is defined as the mean of the maximum spin splitting across all $\mathbf{k}$-points with eigenstates lying within a ±3 eV window around the Fermi level,
\begin{align}
\langle \Delta \rangle = \frac{1}{N} \sum_{k=1}^{N} \max_{n \in W_k} \left| \varepsilon_{n,k}^{\uparrow} - \varepsilon_{n,k}^{\downarrow} \right|,
\end{align}
where $W_k$ is the set of bands at $\mathbf{k}$ within the energy window, and $N$ is the number of contributing $\mathbf{k}$-points.

The volumetric spin-splitting fraction $F_\Delta$ captures the fraction of bands and the portion of the Brillouin zone exhibiting appreciable spin splitting:
\begin{align}
F_\Delta = \sum_k w_k \cdot \Theta\left( \frac{1}{M_k} \sum_{n \in W_k} \left| \varepsilon_{n,k}^{\uparrow} - \varepsilon_{n,k}^{\downarrow} \right| \right),
\end{align}
where $w_k$ is the $\mathbf{k}$-point weight, $M_k$ is the number of bands within the energy window at $\mathbf{k}$, and $\Theta$ is the Heaviside function applied with a small cutoff (e.g., 10 meV) to suppress numerical noise. 

Note that the calculations are performed using the usual scalar-relativistic projector augmented-wave (PAW) pseudopotentials distributed with VASP, but we do not otherwise include spin–orbit coupling (SOC) or other relativistic corrections. This defines the non-relativistic baseline for spin splitting in altermagnets, where the essential effect arises from exchange and crystalline symmetry rather than relativistic interactions. For compounds containing heavy 5d or 5f elements, however, SOC can strongly influence electronic phases \cite{Picozzi2025} and consequently affect the magnitude of the spin splitting. The explicit inclusion of SOC in high-throughput workflows is computationally demanding and, in heavy-element systems, may also lead to convergence difficulties. Hence, in particular our results for heavy-element compounds, which are expected to be more sensitive to these effects, may need further investigation for the contribution from SOC.

\begin{figure*}[t]
  \centering
  \includegraphics[width=\textwidth,trim=18pt 6pt 6pt 6pt,clip]{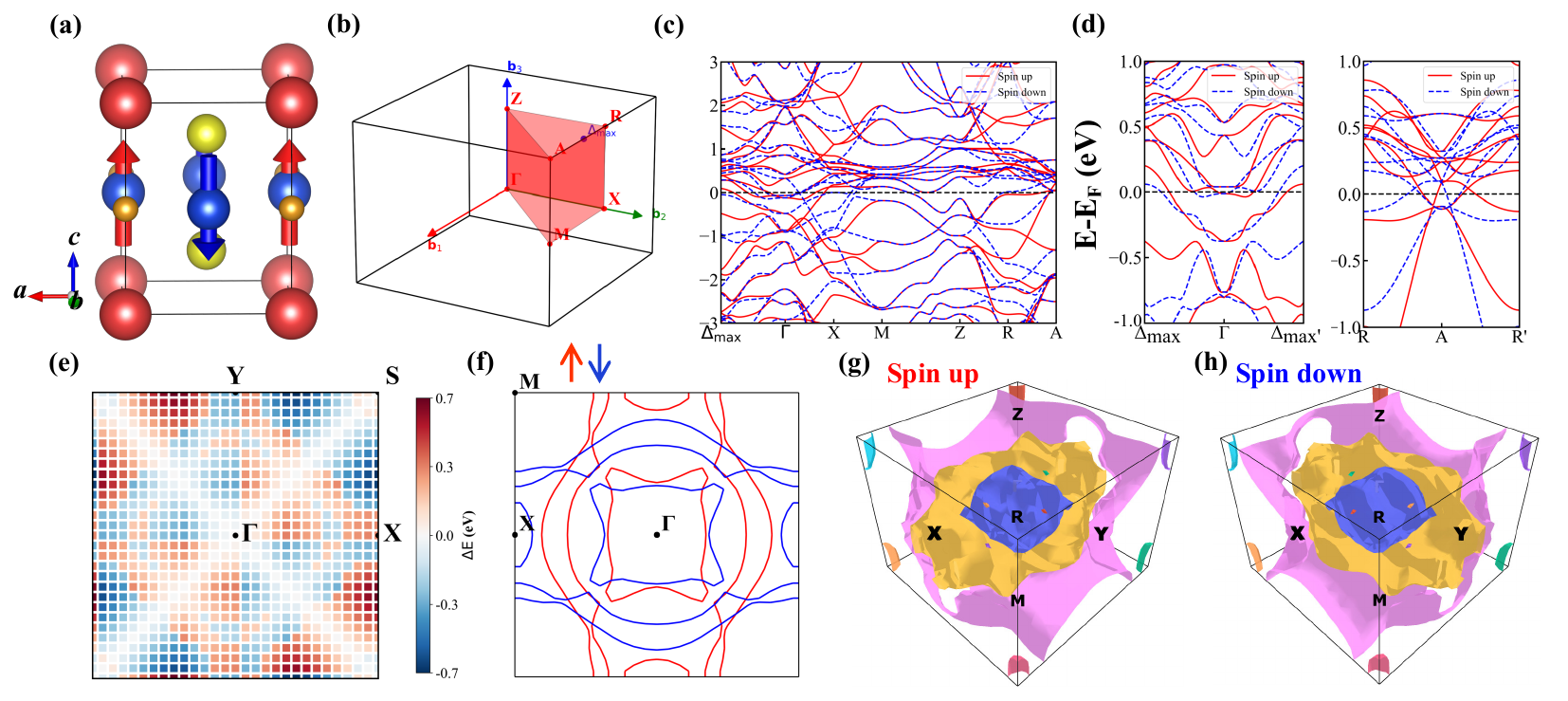}
    \caption{\textbf{Structural, electronic, and spin-resolved properties of UCr$_2$Si$_2$C.} 
\textbf{(a)} Side view of the crystal structure of UCr$_2$Si$_2$C, illustrating the magnetic configuration with arrows representing spin orientations. 
\textbf{(b)} Corresponding Brillouin zone (BZ) indicating key high-symmetry points. 
\textbf{(c)} Spin-polarized electronic band structure along high-symmetry paths, where spin-up and spin-down bands are represented by red and blue dashed lines, respectively. 
\textbf{(d)} Band structures at symmetry-equivalent paths within the BZ, highlighting the inversion of spin-up and spin-down bands due to symmetry operations. 
\textbf{(e)} Two-dimensional contour plot of spin splitting ($\Delta E$) at $k_z = 0.5$, showing the plane with maximal spin splitting. 
\textbf{(f)} The Fermi surface cut at $k_z$$=$ 0, through the $\Gamma$$-$X$-$M plane, illustrating spin-up (red) and spin-down (blue) contributions. 
\textbf{(g--h)} Three-dimensional (3D) spin-resolved Fermi surfaces for \textbf{(g)} spin-up and \textbf{(h)} spin-down electronic states, emphasizing spin-dependent characteristics. Note that the absolute orientation of the collinear spin axis does not affect DFT calculations without spin-orbit coupling, hence, in our pictures the spin directions are shown as computed (in +/- z direction) rather than as reported for the respective structures.}
    \label{fig:fig2}
\end{figure*}

\section{Results and discussion}

Summarized in Table~\ref{tab:materials} are the key properties of the top candidate materials identified through our high-throughput screening, including their maximum, average, and volumetric spin-splitting magnitudes ($\Delta F$), electronic band gaps, Hubbard $U$ parameters, and minimum elemental abundances. Among these, six altermagnets, CrSb, MnTe, CrSe, RuO$_2$, Ca(Al$_2$Fe)$_4$, and UCr$_2$Si$_2$C, exhibit particularly pronounced spin-band splitting. Notably, experimental evidence for altermagnetism in CrSb \cite{reimers2024direct}, MnTe \cite{liu2024chiral}, and RuO$_2$ \cite{fedchenko2024observation} has recently been presented, providing strong validation for the predictive capability of our computational framework. 
In the following sections, we focus on three representative cases: the metallic altermagnets UCr$_2$Si$_2$C and MnNbP, and the semiconducting altermagnet YRuO$_3$, each exhibiting spin-splitting features and serving as compelling examples of the diverse electronic characteristics accessible within the altermagnetic landscape. 

 
\subsection{UCr$_2$Si$_2$C}
UCr$_2$Si$_2$C crystallizes in a tetragonal structure belonging to the paramagnetic space group $P4/mmm$ (No. 123), characterized by lattice parameters $a = b = 3.98\mathrm{\AA}$ and $c = 5.16\mathrm{\AA}$ \cite{lemoine2018unexpected}. The structure is composed of alternating uranium and Cr$_2$Si$_2$C layers stacked along the crystallographic $c$-axis. Within each Cr$_2$Si$_2$C layer, chromium atoms form planar Cr$_2$C networks arranged in a checkerboard motif. Magnetic moments are localized exclusively on the Cr atoms, which occupy two symmetry-inequivalent positions: Cr1 at $(0.0,\,0.5,\,0.5)$ and Cr2 at $(0.5,\,0.0,\,0.5)$. These moments are assumed to align antiparallel, with an estimated magnitude of approximately $0.66 \mu_\mathrm{B}$. 

Under this collinear AFM configuration, UCr$_2$Si$_2$C is associated with the magnetic space group $Pm'm'm$ (No. 47.252, BNS notation), which determine its magnetic symmetry properties. This group comprises eight symmetry elements: four unitary operations that preserve spin namely, the identity $\{1 \,|\, \mathbf{0}\}$, spatial inversion $\{-1 \,|\, \mathbf{0}\}$, a twofold rotation about the $x$-axis $\{2_{100} \,|\, \mathbf{0}\}$, and a mirror reflection across the plane perpendicular to the $x$-axis $\{m_{100} \,|\, \mathbf{0}\}$ as well as four anti-unitary operations combining spatial transformations with time-reversal symmetry. These include the twofold rotations about the $z$-axis combined with time reversal $\{2'_{001} \,|\, \mathbf{0}\}$ and a twofold rotation about the $y$-axis combined with time reversal $\{2'_{010} \,|\, \mathbf{0}\}$, and the time-reversed mirror reflections $\{m'_{001} \,|\, \mathbf{0}\}$ and $\{m'_{010} \,|\, \mathbf{0}\}$. These operations collectively preserve the overall tetragonal crystal symmetry while accommodating a collinear antiferromagnetic configuration with Cr.

Crucially, although both inversion $\mathcal{P}$ $\{-1 \,|\, \mathbf{0}\}$ and anti-unitary time revesal $\mathcal{T} = \{-1' \,|\, \mathbf{0}\}$ symmetries appear in the group, the combined operation $\mathcal{PT} = \{-1' \,|\, \mathbf{0}\}$ does not constitute a symmetry of the system. The absence of $\mathcal{PT}$ symmetry lifts the constraint that would otherwise enforce spin degeneracy, thereby permitting spin splitting even in the absence of spin–orbit coupling. Furthermore, the system lacks the combined spin-rotation and non-primitive lattice translation symmetry $U\tau$, which maps antiparallel magnetic sublattices onto one another via a $180^\circ$ spin rotation coupled with a fractional lattice translation. Such symmetries are characteristic of conventional collinear antiferromagnets and protect spin degeneracy throughout the Brillouin zone. In contrast, the symmetry-inequivalent positioning of the Cr atoms in UCr$_2$Si$_2$C precludes the existence of any $U\tau$ operation. The simultaneous breaking of both $\mathcal{PT}$ and $U\tau$ symmetries thus defines the altermagnetic nature of UCr$_2$Si$_2$C.

Figure \ref{fig:fig2}(c) presents the spin-polarized electronic band structure of UCr$_2$Si$_2$C along high-symmetry paths in the Brillouin zone, as defined in Fig.\ref{fig:fig2}(b). Spin-up and spin-down states are shown in red and blue, respectively. A prominent feature of the band structure is the coexistence of spin-degenerate and spin-split regions across momentum space, hallmarks of altermagnetic behavior. The maximum energy separation occurs near the $\Delta_\text{max}$ point, where the spin splitting reaches approximately 0.72 eV. The average spin splitting across the Brillouin zone is calculated to be 0.31eV, with a volumetric splitting ratio of roughly 47\%. To explicitly identify symmetry-driven spin reversals, we analyzed symmetry-equivalent paths in the reciprocal space, as shown in panel (d). Specifically, along paths such as $\Delta_{\mathrm{max}}$--$\Gamma$--$\Delta'_{\mathrm{max}}$ and $R$--$A$--$R'$, exchanging the reciprocal-space coordinates $k_x \leftrightarrow k_y$ directly leads to inversion of the spin polarization (spin-up states become spin-down and vice versa). This momentum-dependent spin reversal arises naturally from the magnetic symmetry operations of the $Pm'm'm$ space group and it distinguishes altermagnetic systems from conventional ferromagnets or Rashba-type spin–orbit coupled systems.

To further visualize the momentum-resolved spin splitting, we plot the two-dimensional (2D) contour map of the spin splitting magnitude ($\Delta E$) on the $k_z = 0.5$ plane in Fig.~\ref{fig:fig2}(e). This plane includes the maximum splitting point $\Delta_{\text{max}} = (0.2, 0.5, 0.5)$, where the maximum spin splitting occurs. In this map, red and blue regions respectively indicate positive and negative spin splitting between spin-up and spin-down states. The most intense features appear in the vicinity of $\Delta_{\text{max}}$, with symmetric but sign-reversed spin splitting observed on opposite sides of the Brillouin zone. This antisymmetric pattern of spin splitting, i.e., $\Delta E(\mathbf{k}) = -\Delta E(-\mathbf{k})$, is a defining signature of altermagnetic order in this material.

Figure \ref{fig:fig2}(f) shows the spin-resolved Fermi surface cross-sections in the (001) plane. Spin-up and spin-down Fermi contours are shown in red and blue, respectively. The contours are anisotropic and exhibit a clear $90^\circ$ rotation between the spin channels, another hallmark of altermagnetism. To fully capture the spin-dependent topology of the Fermi surface, we plot the three-dimensional spin-resolved Fermi surfaces in Figs.~\ref{fig:fig2}(g) and \ref{fig:fig2}(h) for the spin-up and spin-down channels, respectively. Each plot reveals four distinct Fermi surfaces corresponding to four partially occupied bands crossing the Fermi level. Notably, while the innermost Fermi surface centered at the corners of the Brillouin zone appears nearly identical for both spin channels, the remaining three Fermi sheets (colored yellow, blue, and pink) exhibit clear rotational asymmetry. The spin-down Fermi surfaces are clearly rotated by $90^\circ$ relative to their spin-up counterparts, consistent with the antisymmetric spin texture expected from the symmetry constraints of the magnetic space group. 
Together, these observations, momentum-antisymmetric spin splitting, 90$^\circ$-rotated spin-resolved Fermi surfaces, and symmetry-induced spin channel inversion, provide compelling evidence for the robust altermagnetic character of UCr$_2$Si$_2$C.
\begin{figure*}[t]
  \centering
  \includegraphics[width=\textwidth,trim=18pt 6pt 6pt 6pt,clip]{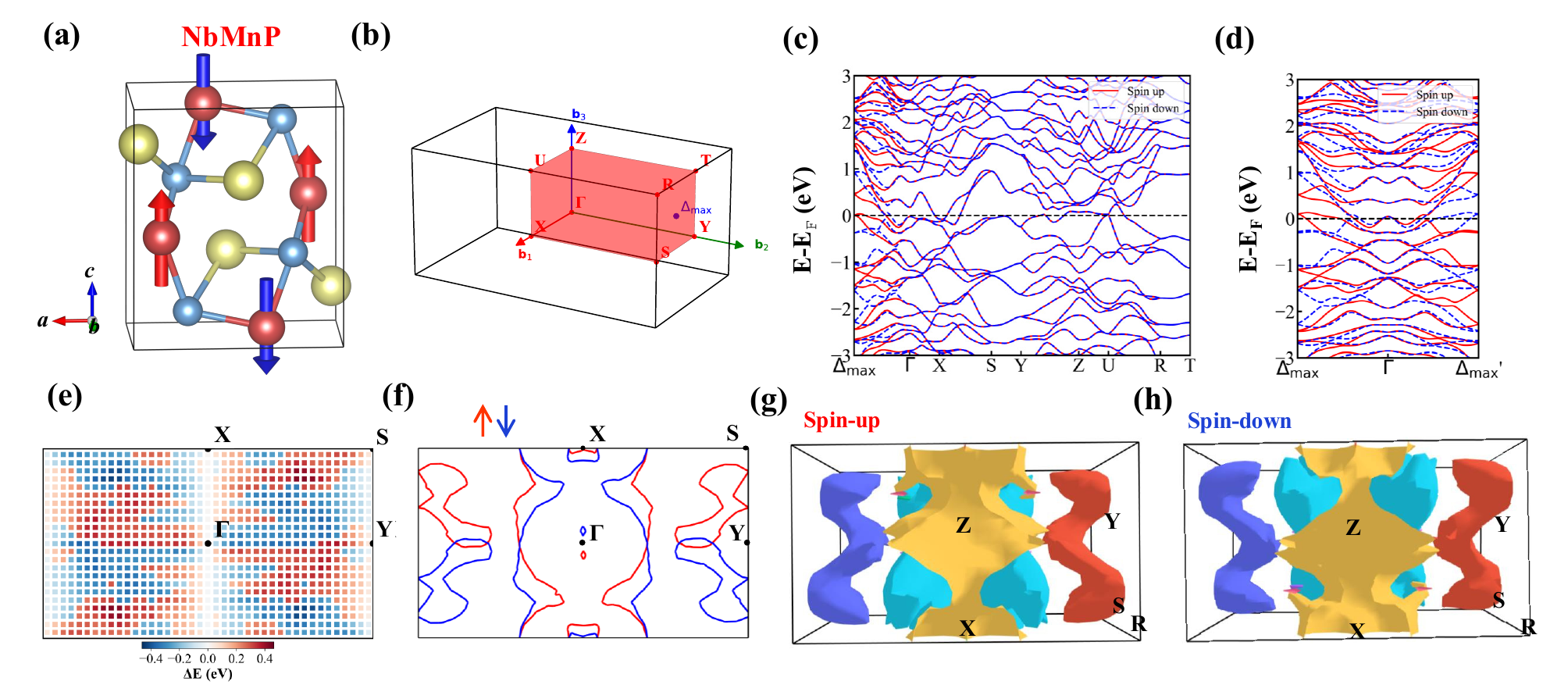}
    \caption{
        \textbf{Structural, electronic, and spin-resolved properties of NbMnP.}
        (a) Side view of the crystal structure of NbMnP, illustrating the AFM configuration with spins aligned along the z-axis on the Mn sites.
        (b) Corresponding BZ indicating key high-symmetry points.
        (c) Spin-polarized electronic band structure along standard high-symmetry paths, with spin-up and spin-down channels represented by red and blue lines, respectively. 
        (d) Spin-resolved band structure along the $\Delta_{max}-\Gamma-\Delta_{max}^{\prime}$ path, highlighting the symmetry-enforced spin reversal. 
        (e) 2D contour map of the spin-splitting magnitude ($\Delta E$) on the $k_z = 0.25$ plane. 
        (f) Spin-resolved Fermi surface cut at $k_z$=0.25 plane. These contours are anisotropic and exhibit a $180^\circ$ rotation between the spin channels.
        (g) and (h) 3D spin-resolved Fermi surfaces for the spin-up and spin-down channels, respectively, showing complementary shapes related by $k_x \leftrightarrow -k_x$ and a spin flip. 
    }
    \label{fig:fig3}
\end{figure*}

\subsection{NbMnP}

NbMnP crystallizes in the TiNiSi$-$type orthorhombic structure belonging to the paramagnetic space group \textit{Pnma} (No. 62), characterized by lattice parameters $a = 6.182\,\mathrm{\AA}$, $b = 3.557\,\mathrm{\AA}$, and $c = 7.219\,\mathrm{\AA}$ \cite{matsuda2021noncollinear}. Its crystal structure comprises zigzag chains of Mn atoms aligned along the $b$-axis, embedded within a network formed by Nb and P atoms. In this study, we consider a collinear AFM configuration as depicted in Fig.~\ref{fig:fig3}(a).

The magnetic symmetry properties of this AFM state are governed by magnetic space group $Pn'm'a'$ (No. 62.449, BNS notation), a type-III magnetic subgroup of the paramagnetic parent group Pnma (No. 62). This magnetic group incorporates eight symmetry elements: four unitary operations preserving spin specifically, the identity $\{1\mid\mathbf{0}\}$, spatial inversion $\{-1\mid\mathbf{0}\}$, mirror reflections across the $yz$-plane $\{m_{100} \mid \tfrac{1}{2}, \tfrac{1}{2}, \tfrac{1}{2}\}$, and the $xz$-plane $\{m_{010} \mid \tfrac{1}{2}, \tfrac{1}{2}, \tfrac{1}{2}\}$; and four anti-unitary operations combining spatial transformations with time reversal, including time-reversed inversion $\{-1' \mid \tfrac{1}{2}, 0, \tfrac{1}{2}\}$, a time-reversed mirror reflection across the $xy$-plane $\{m'_{001} \mid \tfrac{1}{2}, 0, \tfrac{1}{2}\}$, spin-flip mirror operations $\{m'_{100} \mid 0, \tfrac{1}{2}, 0\}$, and $\{m'_{010} \mid 0, \tfrac{1}{2}, 0\}$. These symmetry operations accommodate the collinear AFM order with alternating Mn spins. Critically, both $\mathcal{PT}$ and $U_\tau$ symmetries are absent which implies that no symmetry exists to map the antiparallel Mn sublattices onto one another, thereby lifting the constraints that would typically enforce spin degeneracy in conventional antiferromagnets. 

The spin-polarized electronic band structure along standard high-symmetry paths is presented in Fig.~\ref{fig:fig3}(c), with spin-up and spin-down channels shown in red and blue, respectively. Pronounced spin splitting regions coexist with segments of near degeneracy. The maximum spin splitting, approximately 0.46\,eV, occurs near the $\Delta_\mathrm{max}$ point, whereas the BZ-averaged splitting is about 0.28 eV with a volumetric splitting ratio of 35\%. To illustrate symmetry-enforced spin reversal explicitly, Fig.~\ref{fig:fig3}(d) displays the spin-resolved bands along the path $\Delta_{\mathrm{max}}$--$\Gamma$--$\Delta_{\mathrm{max}}'$, demonstrating symmetric spin-band exchange under the transformation $k_x \rightarrow -k_x$ across $\Gamma$.

Figure \ref{fig:fig3}(e) shows the spin-splitting magnitude $\Delta E$ within the $k_z = 0.25$ plane, revealing clearly sign-reversed splitting distributions across opposite sides of the BZ. Additionally, Fig.~\ref{fig:fig3}(f) presents the spin-resolved Fermi surface cross-section at $k_z = 0.05$, exhibiting distinctly anisotropic contours and a clear 180$^\circ$ rotation between the spin channels, consistent with $k_x \leftrightarrow -k_x$ symmetry. The 3D spin-resolved Fermi surfaces shown in Figs.~\ref{fig:fig3}(g) and \ref{fig:fig3}(h) for spin-up and spin-down channels, respectively, further highlight this 180$^\circ$ spin-channel rotation. Taken together, these observations robustly confirm the altermagnetic character of NbMnP.

\begin{figure*}[t]
  \centering
  \includegraphics[width=\textwidth,trim=18pt 6pt 6pt 6pt,clip]{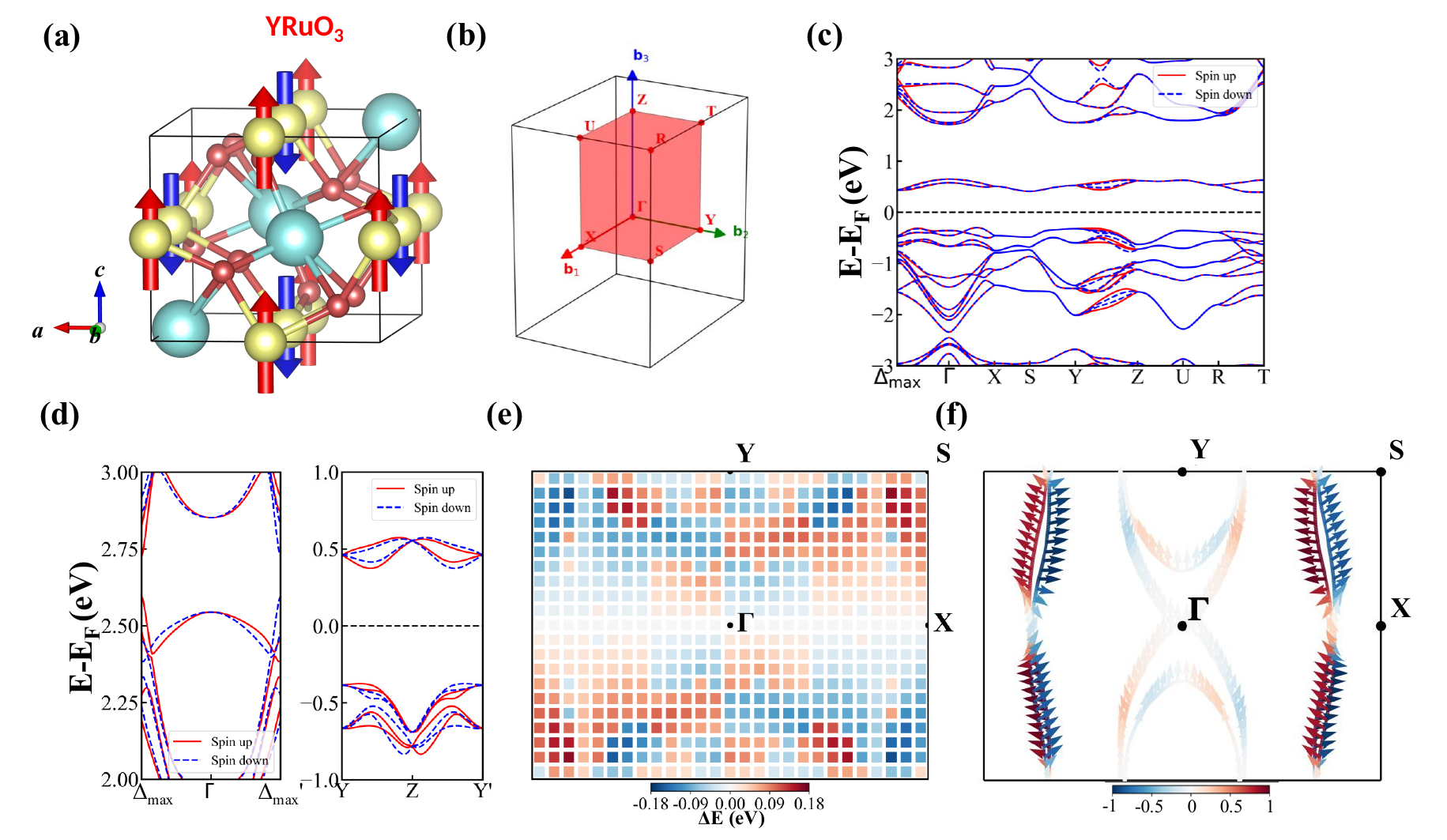}
    \caption{\textbf{Structural, electronic, and spin-resolved properties of YRuO$_3$}. (a) Unit cell of YRuO$_3$, illustrating the atomic arrangement and the AFM spin configuration. Large light green spheres represent Y atoms, small red spheres represent O atoms, and yellow spheres represent Ru atoms. (b) Bulk Brillouin zone of YRuO$_3$ with high-symmetry points labeled. (c) Spin-polarized electronic band structure of YRuO$_3$ along high-symmetry directions. Red solid lines indicate spin-up bands, and blue dashed lines represent spin-down bands. (d) Detailed view of the spin-polarized band structure along two symmetry-equivalent paths within the BZ, highlighting the spin inversion features. (e) 2D contour plot of the spin splitting $\Delta$E in the k$_z$=0.4 plane, identifying regions of maximal spin-momentum locking. (f) Spin texture visualized in the 001 plane, showing the orientation and magnitude of the spin polarization for selected states.}
    \label{fig:fig4}
\end{figure*}

\subsection{YRuO$_3$}
YRuO$_3$ crystallizes in the orthorhombic magnetic space group $Pn'ma'$ (No.\ 62.448, BNS notation), derived from the paramagnetic parent group Pnma (No.\ 62) \cite{ji2020yruo}. The unit cell adopts lattice parameters $a=5.839\,\text{\AA}$, $b=7.523\,\text{\AA}$, $c=5.192\,\text{\AA}$. The crystal structure can be described as a tilted perovskite framework composed of corner-sharing RuO$_6$ octahedra, with Y atoms occupying the interstitial positions. The magnetic ground state is G-type antiferromagnetic, with antiparallel spin alignment between neighboring Ru$^{3+}$ sites along all three crystallographic directions. The spin moments are localized on the Ru sublattices, whereas Y and O atoms remain magnetically inactive. Addtionally, it includes standard unitary operations, such as inversion, mirror planes, and twofold rotations, along with anti-unitary operations that combine time reversal with rotations and mirrors (e.g., $2'_{001}$, $m'_{100}$, $m'_{001}$). The resulting AFM structure consists of two symmetry-inequivalent Ru sublattices with antiparallel spin orientations. This arrangement prevents any symmetry operation from mapping one sublattice onto the other via inversion--time-reversal ($\mathcal{PT}$) or spin-rotation--translation ($\mathcal{U}\tau$). This dual symmetry breaking allows for spin splitting even without spin--orbit coupling, which is what defines the altermagnetic nature of YRuO$_3$.

Figure \ref{fig:fig4}(b) shows the bulk Brillouin zone BZ YRuO$_3$ with the high-symmetry points labeled according to the Pnma convention. The spin-polarized electronic band structure, computed along these high-symmetry paths, is presented in Fig.~\ref{fig:fig4}(c). A key feature is the coexistence of spin-degenerate and spin-split bands. The magnitude of spin splitting varies across momentum space, reaching its maximum near the $\Delta_{\mathrm{max}}$ point, where the energy separation between spin-up and spin-down bands is approximately 0.12 eV.

To visualize the symmetry-driven spin inversion, Fig.~\ref{fig:fig4}(d) compares the band dispersions along two momentum paths that are equivalent under the transformation $k_x \rightarrow -k_x$. Along these symmetry-related paths, the spin polarization reverses: spin-up states on the $\Delta_{\mathrm{max}}$/Y side correspond to spin-down states on the $-\Delta_{\mathrm{max}}$/$-$Y side, reflecting the momentum-antisymmetric spin splitting enforced by the \( Pn'ma' \) magnetic group. Figure~\ref{fig:fig4}(e) further highlights this feature through a 2D contour map of \( \Delta E \) in the \( k_z = 0.1 \) plane, where the splitting is maximal around \( \Delta_{\text{max}} \). Figure~\ref{fig:fig4}(f) illustrates the spin texture in the (001) plane from a SOC (noncollinear) calculation, showing the in-plane orientation of the spin expectation value for states near the Fermi level. The spin-up and spin-down components exhibit opposite orientations across symmetry-related k-points, indicating a momentum-dependent spin reversal that extends across the Brillouin zone rather than being localized to a specific path or point.

\section{Identification in 2D databases}
To identify bulk altermagnetic compounds with chemically equivalent 2D counterparts, we compared the reduced formulas and ICSD IDs of
all 188 bulk AM candidates with entries in two major 2D materials databases: the Computational 2D Materials Database (C2DB)~\cite{haastrup2018c2db} and the AiiDA 2D repository~\cite{pizzi2016aiida}.  We only retain entries for which the ICSD ID of the bulk parent coincides with that of the corresponding 2D record.
In Table~\ref{tab:2d_matches}, we collect all bulk AM materials whose reduced formula appears in at least one of the 2D databases and indicate whether a match was found in C2DB, in AiiDA, or in both. These databases contain both experimentally synthesized and theoretically predicted monolayer compounds. 


\begin{table}[t]
\caption{Bulk altermagnetic materials with an existing 2D compound of the same reduced chemical formula in the C2DB or AiiDA databases. Multiple MCIF entries with identical formulas correspond to distinct MAGNDATA models.}
\label{tab:2d_matches}
\centering

\begin{tabular*}{\columnwidth}{@{\extracolsep{\fill}} l l c c }
\hline\hline
IDs        & Chemical formula      & C2DB & AiiDA \\
\hline
0.15            & MnF$_2$      & Yes  & --    \\
0.178, 0.334    & CoF$_2$      & Yes  & --    \\
0.335, 0.581, 1.0.33 & FeF$_3$ & Yes  & --    \\
0.36            & NiF$_2$      & Yes  & --    \\
0.607           & RuO$_2$      & Yes  & --    \\
0.800           & MnTe         & Yes  & --    \\
1.0.47, 1.0.48  & MnSe$_2$     & Yes  & Yes   \\
2.35            & CrSe         & Yes  & --    \\
0.65, 0.66      & Fe$_2$O$_3$  & --   & Yes   \\
\hline\hline
\end{tabular*}
\end{table}

The list of bulk altermagnets with a corresponding two-dimensional compound is rather short.
Indeed, the majority of compounds appearing in Table~\ref{tab:2d_matches} contain transition–metal elements with partially filled $d$ shells together with halogens (F, Cl) or chalcogens (S, Se, Te). The later help stabilizing layered or quasi-layered chemistries, while the former provide the local moments that form the basis of the magnetism. 

Note that the first column of the table shows that some chemical formulas correspond to multiple MAGNDATA entries. The reason is that different experimental studies report distinct magnetic configurations, magnetic subgroup assignments, or slightly different crystallographic refinements. These duplicated entries correspond to separate magnetic models of the same material, not different compounds, and therefore belong match with a single two-dimensional material.  

Table~\ref{tab:2d_matches} identifies candidate bulk altermagnets which exist in 2D, but does not say that these materials are also altermagnetic as a monolayer. This question has been addressed by S\o{}dequist and Olsen~\cite{sodequist24}, who identified 7 altermagnetic monolayer candidates starting from C2DB. Crucially, they take into account the magnetic anisotropy coming from spin-orbit that is necessary to get a magnetically ordered phase in 2d, in accordance with the Mermin-Wagner theorem. Starting from C2DB, they identified which monolayers can be altermagnets, also taking into account spin-orbit coupling. The spin-orbit coupling introduces magnetic anisotropy that makes it possible to stabilize magnetism in 2d while satisfying the Mermin-Wagner theorem. Their 7 altermagnetic monolayer candidates similarly contain mostly transition metals terminated by halogens or chalcogens, although they also find two more complex compounds. Interestingly, there is zero overlap between Table~\ref{tab:2d_matches} and the list in S\o{}dequist and Olsen.


\section{Comparison with other studies}
\label{sec:comparison}
The results reported here can be discussed in relation to two prior screenings of MAGNDATA. The high-throughput classification by Wan \textit{et al.}\ relied on the absence of the antiunitary symmetries $\mathcal{PT}$ (inversion plus time reversal) and $t$$\mathcal{T}$ (translation plus time reversal)~\cite{wan2024high}. Moreover, they extended their candidate pool to include ferrimagnetic systems by considering their SOC-off configuration, resulting in 336 potential altermagnets. Guo \textit{et al.} also focused on systems with $\mathcal{PT}$ symmetry~\cite{Guo2023}. They screened stoichiometric compounds with fully compensated collinear magnetic order without ferromagnetic component while avoiding some complications, including systems with magnetic order from rare-earth atoms, large unit cells, and DFT calculations difficult to converge, which resulted in 62 materials.

Our study employs the \texttt{amcheck} algorithm, which tests the same symmetry condition (absence of $\mathcal{PT}$ and $t$$\mathcal{T}$) to identify potential altermagnetic materials. However, \texttt{amcheck} strictly excludes ferrimagnetic states since their sublattices are completely not symmetry-related and are instead classified as “Luttinger ferrimagnets” that exhibit spin splitting but lack the required momentum-odd, nodal spin texture over the Brillouin zone. For example, Cr\(_2\)S\(_3\), listed as “interesting’’ in Ref.~\cite{wan2024high}, shows spin splitting in the spin–polarized DFT consistent with ferrimagnetic behavior, yet lacks the nodal spin texture and is therefore not classified as altermagnetic here. On the other hand, we find a collinear AFM with Kramers–degenerate bands for SrMnSb\(_2\), and the small splittings of a few meV that appear at high energy are most likely numerical artifacts and not indicative of altermagnetism. These distinctions likely account for the smaller number of altermagnets identified in our study compared to Ref.~\cite{wan2024high}, while remaining fully consistent with the symmetry–based definition of altermagnetism~\cite{vsmejkal2022beyond,vsmejkal2022emerging}. The 59 non-metallic and 3 metallic compounds identified in Ref.~\cite{Guo2023} all appear in our list of 157 collinear entries with the right symmetries for altermagnetism (see table S2 in the supplementary materials), however, the criterion of a spin splitting threshold of 26 meV for our final list removes 10 of them.

Furthermore, we have considered a set of collinear versions of materials reported as noncollinear. This extended set of materials along with their computed spin splittings describe alternative spin configurations on otherwise stable crystal structures.  They can be a valuable resource to aid the general understanding of altermagnetism. Some of them are arguably relevant due to them being very nearly collinear, shown in our analysis by the angle between the spin axes. We also speculate that for some of the ones with larger such angles, there could exist ways in practice to invoke a more aligned moments to give the altermagnetic configuration, via, e.g., external fields, stress/strain, or similar mechanisms.

\begin{table*}[t]
\caption{\textbf{Top altermagnet candidate materials with respect to their spin-splitting.}
Space group, max angle between spins for materials that were originally noncollinear ($\arccos (|S_i \cdot S_j|/|S_i||S_j|)$), spin-splitting metrics, band gap, wave classification, Hubbard $U$, and minimum elemental abundance. For more information on symmetry data, refer to the supplementary materials.}
\label{tab:materials}
\centering
\begin{tabular}{l c c c c c c c l c}
\hline\hline
Material & Space Group & $\frac{|S_i \cdot S_j|}{|S_i||S_j|}$ & $F_\Delta$ (\%) & Max.\ SS (eV) & Avg.\ SS (eV) & Bandgap (eV) & Wave & $U$ Values & Min Abund. \\
\hline
\ch{CrSb} & P6$_3$/mmc & 0.0$^\circ$ & 34.375 & 1.872 & 0.763 & 0 & g & Cr: 3.5 & $2\times10^{3}$ \\
\ch{MnTe} & P6$_3$/mmc & 0.0$^\circ$ & 34.375 & 0.923 & 0.449 & 0.7637 & g & Mn: 4.0 & $1\times10^{1}$ \\
\ch{RuO2} & P4$_2$/mnm & 0.0$^\circ$ & 25.000 & 0.865 & 0.351 & 0 & d & Ru: 2.0 & $1\times10^{1}$ \\
\ch{CrSe} & P6$_3$/mmc & 57.79$^\circ$ & 35.938 & 0.800 & 0.489 & 0 & g & Cr: 3.5 & $5\times10^{2}$ \\
\ch{UCr2Si2C} & P4/mmm & 0.0$^\circ$ & 46.875 & 0.719 & 0.312 & 0 & d & Cr: 3.5 & $3\times10^{4}$ \\
\ch{Ca(Al2Fe)4} & I4/mmm & 8.41$^\circ$ & 37.500 & 0.628 & 0.374 & 0 & d & Fe: 5.3 & $4\times10^{8}$ \\
\ch{CoF2} & P4$_2$/mnm & 0.0$^\circ$ & 16.667 & 0.479 & 0.223 & 3.3056 & d & Co: 3.32 & $2\times10^{5}$ \\
\ch{MnNbP} & Pnma & 86.65$^\circ$ & 35.000 & 0.457 & 0.276 & 0 & d & Mn: 4.0;\ Nb: 1.5 & $2\times10^{5}$ \\
\ch{NiF2} & P4$_2$/mnm & 1.72$^\circ$ & 27.083 & 0.392 & 0.180 & 5.2675 & d & Ni: 6.4 & $8\times10^{5}$ \\
\ch{LiFe2F6} & P4$_2$/mnm & 0.0$^\circ$ & 12.500 & 0.319 & 0.159 & 0.1679 & d & Fe: 5.3 & $2\times10^{5}$ \\
\ch{Mn(C2N3)2} & Pnnm & 0.0$^\circ$ & 20.000 & 0.305 & 0.184 & 4.0756 & d & Mn: 4.0 & $2\times10^{5}$ \\
\ch{MnGeN2} & Pna2$_1$ & 0.0$^\circ$ & 14.583 & 0.305 & 0.145 & 1.5162 & d & Mn: 4.0 & $2\times10^{4}$ \\
\ch{FeSO4F} & C2/c & 0.0$^\circ$ & 28.906 & 0.301 & 0.158 & 2.4510 & d & Fe: 5.3 & $4\times10^{6}$ \\
\ch{NiCO3} & R-3c & 0.0$^\circ$ & 20.833 & 0.301 & 0.167 & 4.1969 & g & Ni: 6.4 & $8\times10^{5}$ \\
\ch{MnF2} & P4$_2$/mnm & 0.0$^\circ$ & 18.750 & 0.294 & 0.156 & 3.5754 & d & Mn: 4.0 & $6\times10^{6}$ \\
\ch{CrVO4} & Cmcm & 0.0$^\circ$ & 28.571 & 0.292 & 0.158 & 2.8766 & d & Cr: 3.5;\ V: 3.5 & $1\times10^{6}$ \\
\ch{Sr4Fe4O11} & Cmmm & 0.0$^\circ$ & 29.808 & 0.289 & 0.150 & 0.1756 & d & Fe: 5.3 & $4\times10^{6}$ \\
\ch{La2CoIrO6} & P2$_1$/c & 89.27$^\circ$ & 24.107 & 0.272 & 0.121 & 0 & d & La: 6.0;\ Co: 3.32;\ Ir: 2.0 & $1\times10^{1}$ \\
\ch{TbFeO3} & Pnma & 0.0$^\circ$ & 26.786 & 0.271 & 0.133 & 2.8257 & d & Tb: 6.7;\ Fe: 5.3 & $1\times10^{4}$ \\
\ch{Fe2WO6} & Pbcn & 37.39$^\circ$ & 29.688 & 0.270 & 0.173 & 1.4648 & d & Fe: 5.3 & $1\times10^{4}$ \\
\ch{Er2Ru2O7} & Fd-3m & 0.0$^\circ$ & 20.312 & 0.268 & 0.266 & 0.4121 & d & Er: 6.0;\ Ru: 2.0 & $1\times10^{1}$ \\
\ch{FeBO3} & R-3c & 0.0$^\circ$ & 29.688 & 0.262 & 0.177 & 2.6602 & g & Fe: 5.3 & $1\times10^{5}$ \\
\ch{DyFeO3} & Pnma & 33.4$^\circ$ & 26.786 & 0.251 & 0.133 & 2.8605 & d & Dy: 6.7;\ Fe: 5.3 & $5\times10^{4}$ \\
\ch{HoFeO3} & Pnma & 90.0$^\circ$ & 25.000 & 0.250 & 0.125 & 2.7346 & d & Ho: 6.0;\ Fe: 5.3 & $1\times10^{4}$ \\
\ch{TlCrO3} & Pnma & 0.0$^\circ$ & 24.107 & 0.246 & 0.146 & 0.6254 & d & Cr: 3.5 & $8\times10^{3}$ \\
\hline\hline
\end{tabular}
\end{table*}

\section{Conclusion}
In this work, we have established a comprehensive and symmetry-aware high-throughput framework for the discovery and characterization of altermagnetic materials. We systematically screened more than 2,000 experimentally verified magnetic structures from the MAGNDATA database using a two-stage workflow. This approach, built on the magnetic symmetry diagnostic tool amcheck and the High-Throughput Toolkit, identified 180 candidate materials exhibiting significant momentum-dependent spin splitting. These include both metallic and semiconducting systems, many of which had not been previously reported as altermagnets. Notably, our approach recovers experimentally confirmed compounds such as CrSb, MnTe, and RuO$_2$, and highlights previously overlooked materials like UCr$_2$Si$_2$C, MnNbP, and YRuO$_3$ that display robust symmetry-protected spin splitting even in the absence of spin–orbit coupling. Furthermore, by cross-referencing our bulk AM candidates with the C2DB and AiiDA 2D repository, we identify nine materials with chemically equivalent 2D counterparts. 

Crucially, our momentum-resolved analysis reveals that spin splitting varies strongly across the Brillouin zone, often away from high-symmetry paths, providing actionable insights for future spin and angle-resolved photoemission spectroscopy (ARPES) measurements. The open-access database generated through this effort offers a valuable resource for guiding experimental exploration and materials design in spintronics. More broadly, this study establishes a scalable and transferable blueprint for accelerating the discovery of unconventional magnetic phases and paves the way for targeted development of altermagnetic materials with desirable symmetry and transport properties.

\section{Acknowledgements}
This work was partially supported by the Wallenberg Initiative Materials Science for Sustainability (WISE)
funded by the Knut and Alice Wallenberg Foundation.
R.A. acknowledges financial support from the Swedish e-Science Research Centre (SeRC).
EvL acknowledges support from the Swedish Research Council (Vetenskapsrådet, VR) under grant 2022-03090, from the Royal Physiographic Society in Lund and by eSSENCE, a strategic research area for e-Science, grant number eSSENCE@LU 9:1. J.A.L thank the Kempe-stiftelserna, Sweden and Swedish Research Council under grant no. 2023-03894 for financial support. B.M. acknowledge the financial support of Olle Engkvists stiftelse, project 207-0582. The computations were enabled by resources provided by the National Academic Infrastructure for Supercomputing in Sweden (NAISS), partially funded by the Swedish Research Council through grant agreement no. 2022-06725. 

\appendix 

\section{Computational Details}

Density functional theory (DFT) calculations were carried out using the Vienna \textit{Ab initio} Simulation Package (VASP) \cite{vasp, paw2}, employing the projector augmented-wave (PAW) method and the generalized gradient approximation (GGA) with the Perdew–Burke–Ernzerhof (PBE) functional to describe exchange-correlation effects \cite{paw1,perdew1996generalized}. Electron correlation was treated using the rotationally invariant DFT+$U$ approach \cite{anisimov1991band}, with the effective Hubbard $U_\text{eff}$ parameters taken from Pymatgen (Materials Project) \cite{jain2013} and supplemented by Wang \textit{et al.} \cite{wang2006} where needed. The magnitude of the spin splitting depends on the value of $U$ via the magnitude of the magnetic moments~\cite{Roig24}, whereas the structure of the splitting in the Brillouin Zone is expected to be weakly dependent on the value of $U$. Consistent with this, recalculations with $U$ scaled by $\pm10\%$ for all metallic candidates show that the altermagnetic classification is unchanged and sizable splittings persist; the variations are mainly quantitative and not strictly monotonic, with only Co$_2$Mo$_3$O$_8$ turning insulating at $U{+}10\%$ (Table~S4).

For all simulations, the plane-wave energy cutoff was automatically set to 1.3 times the maximum ENMAX value among the elements present in the system, following VASP's recommended settings and the electronic self-consistency loop was converged to within $10^{-6}$ eV. K-point meshes for SCF calculations were automatically generated by Pymatgen, targeting a density of approximately 1000 k-points per reciprocal atom. For NSCF calculations, the k-point mesh was manually doubled along each reciprocal axis to ensure accurate Brillouin zone sampling for spin-splitting analysis.  Although canted magnetic configurations occur in many materials, for computational efficiency we approximated their magnetic structures using collinear spin arrangements to identify signatures of altermagnetism.

High-throughput calculations were managed using \textit{httk} \cite{httk}. High-symmetry $\mathbf{k}$-paths were constructed using Pymatgen \cite{ong2013python}, and magnetic space groups were identified using the Spglib library \cite{spglib, spglibv2}. Fermi surfaces and spin textures were visualized using the \texttt{IFermi} package \cite{ganose2021ifermi}.

The elemental abundance values, expressed in parts per million (ppm) with respect to the Earth's crust, were obtained using the \texttt{mendeleev} \cite{mendeleev2014} Python library and are reported in the table.


%


\clearpage
\onecolumngrid

\setcounter{page}{1}
\setcounter{table}{0}
\setcounter{figure}{0}
\setcounter{equation}{0}
\setcounter{section}{0}

\renewcommand{\thepage}{S\arabic{page}}
\makeatletter
\renewcommand{\thetable}{S\arabic{table}}
\renewcommand{\thefigure}{S\arabic{figure}}
\renewcommand{\theequation}{S\arabic{equation}}
\renewcommand{\thesection}{S\arabic{section}}
\renewcommand{\tablename}{Table}
\makeatother

\thispagestyle{empty}

\begin{center}
{\large\bfseries Supplemental Material for\par}
\vspace{0.4em}
{\large\itshape High-Throughput Quantification of Altermagnetic Band Splitting\par}
\vspace{1em}

Ali Sufyan,$^{1,2,3}$ Brahim Marfoua,$^{4}$ J. Andreas Larsson,$^{3,5}$ Erik van Loon,$^{1,2}$ and Rickard Armiento$^{4}$

\vspace{0.8em}

{\small
$^{1}$NanoLund and Division of Mathematical Physics, Department of Physics, Lund University, SE-221 00 Lund, Sweden\\
$^{2}$Wallenberg Initiative Materials Science for Sustainability, Department of Physics, Lund University, SE-221 00 Lund, Sweden\\
$^{3}$Applied Physics, Division of Materials Science, Department of Engineering Sciences and Mathematics, Lule\aa\ University of Technology, SE-971 87 Lule\aa, Sweden\\
$^{4}$Department of Physics, Chemistry and Biology, Link\"oping University, SE-581 83 Link\"oping, Sweden\\
$^{5}$Wallenberg Initiative Materials Science for Sustainability, Lule\aa\ University of Technology, SE-971 87 Lule\aa, Sweden
}

\vspace{0.5em}

\email{ erik.van_loon@fysik.lu.se, rickard.armiento@liu.se}
\end{center}

\vspace{1em}

\clearpage

\includepdf[pages=1]{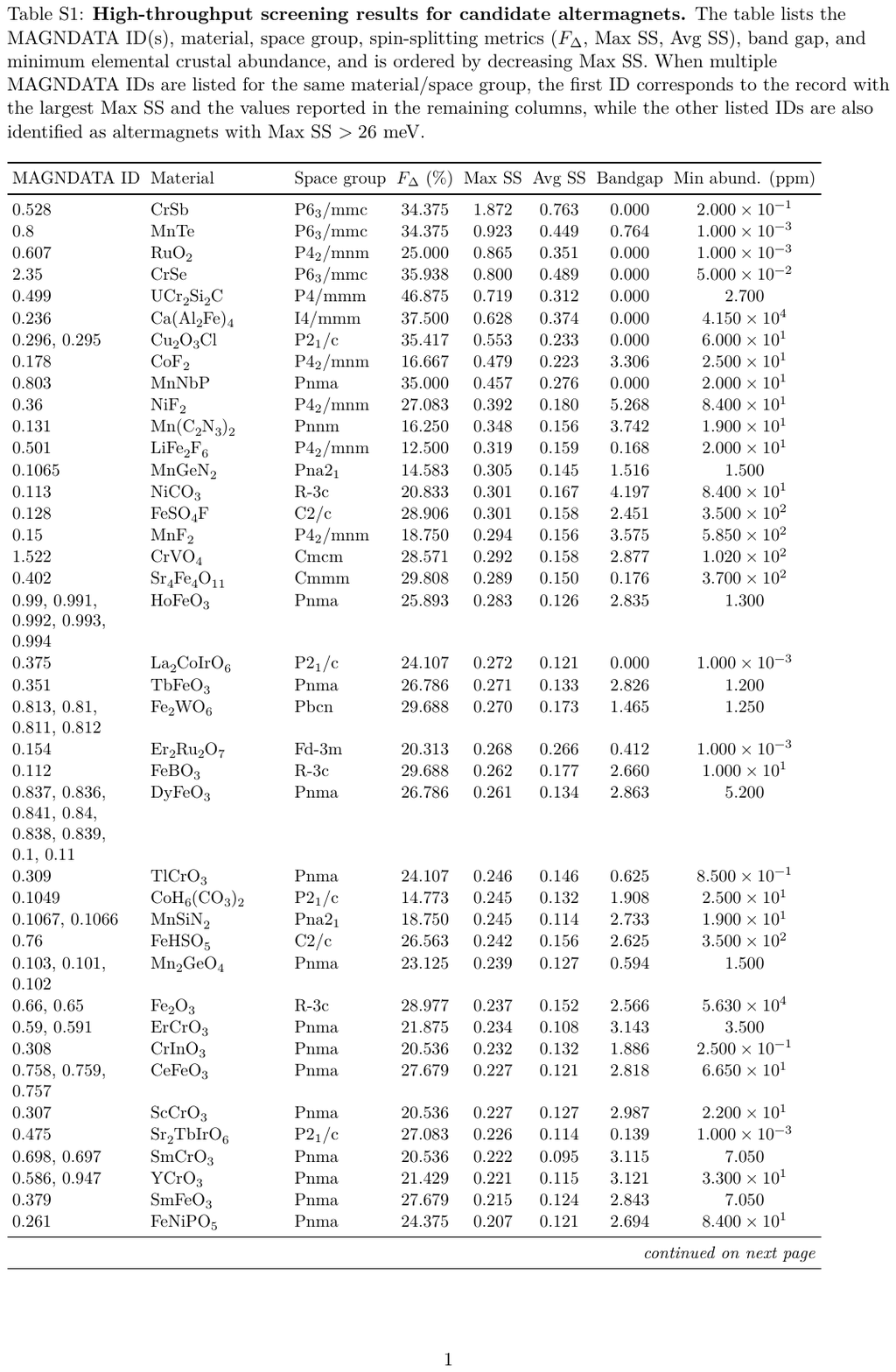}
\includepdf[pages=2]{TableS1.pdf}
\includepdf[pages=3]{TableS1.pdf}
\includepdf[pages=4]{TableS1.pdf}

\clearpage
\includepdf[pages=1]{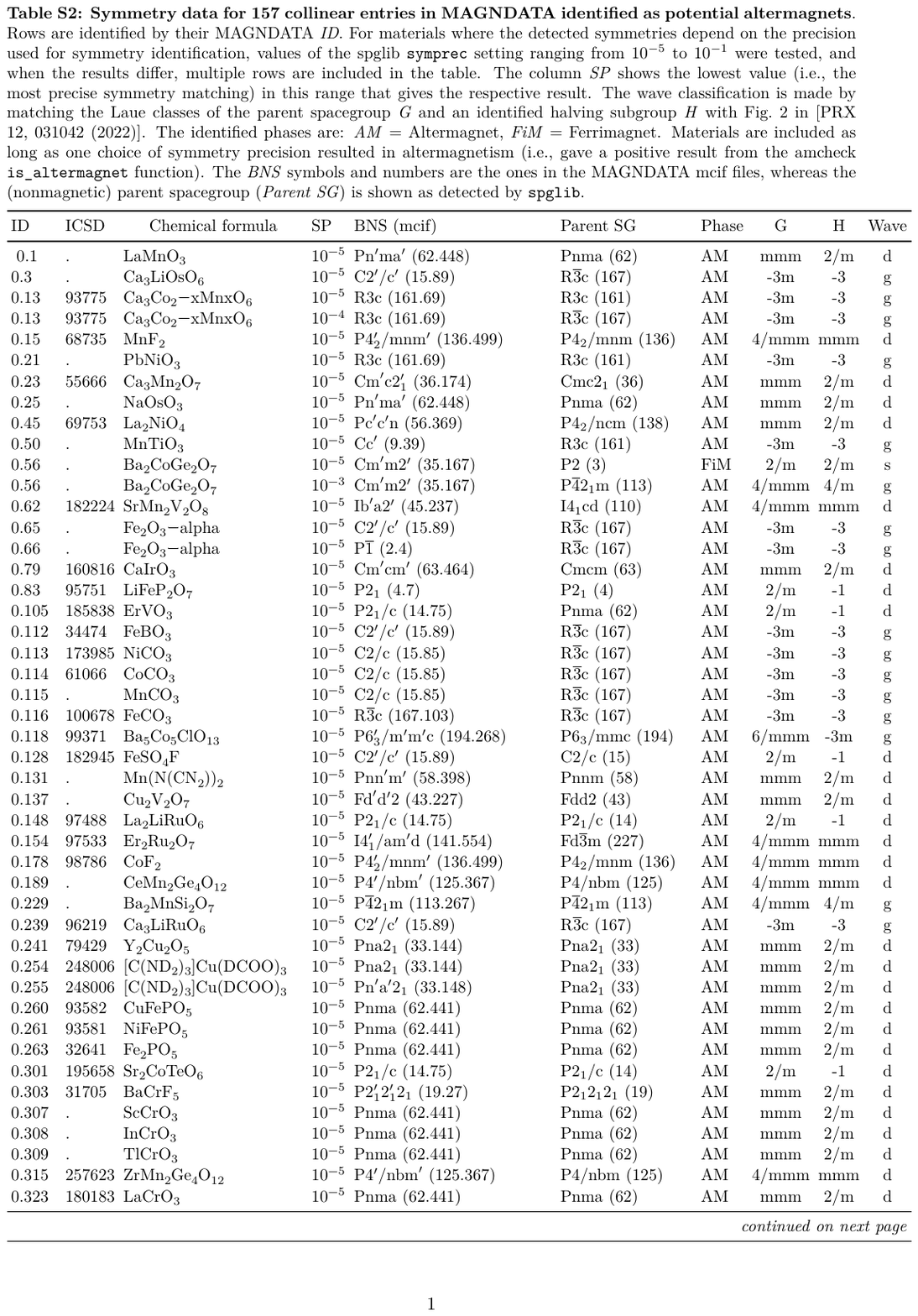}
\includepdf[pages=2]{TableS2.pdf}
\includepdf[pages=3]{TableS2.pdf}
\includepdf[pages=4]{TableS2.pdf}

\clearpage
\includepdf[pages=1]{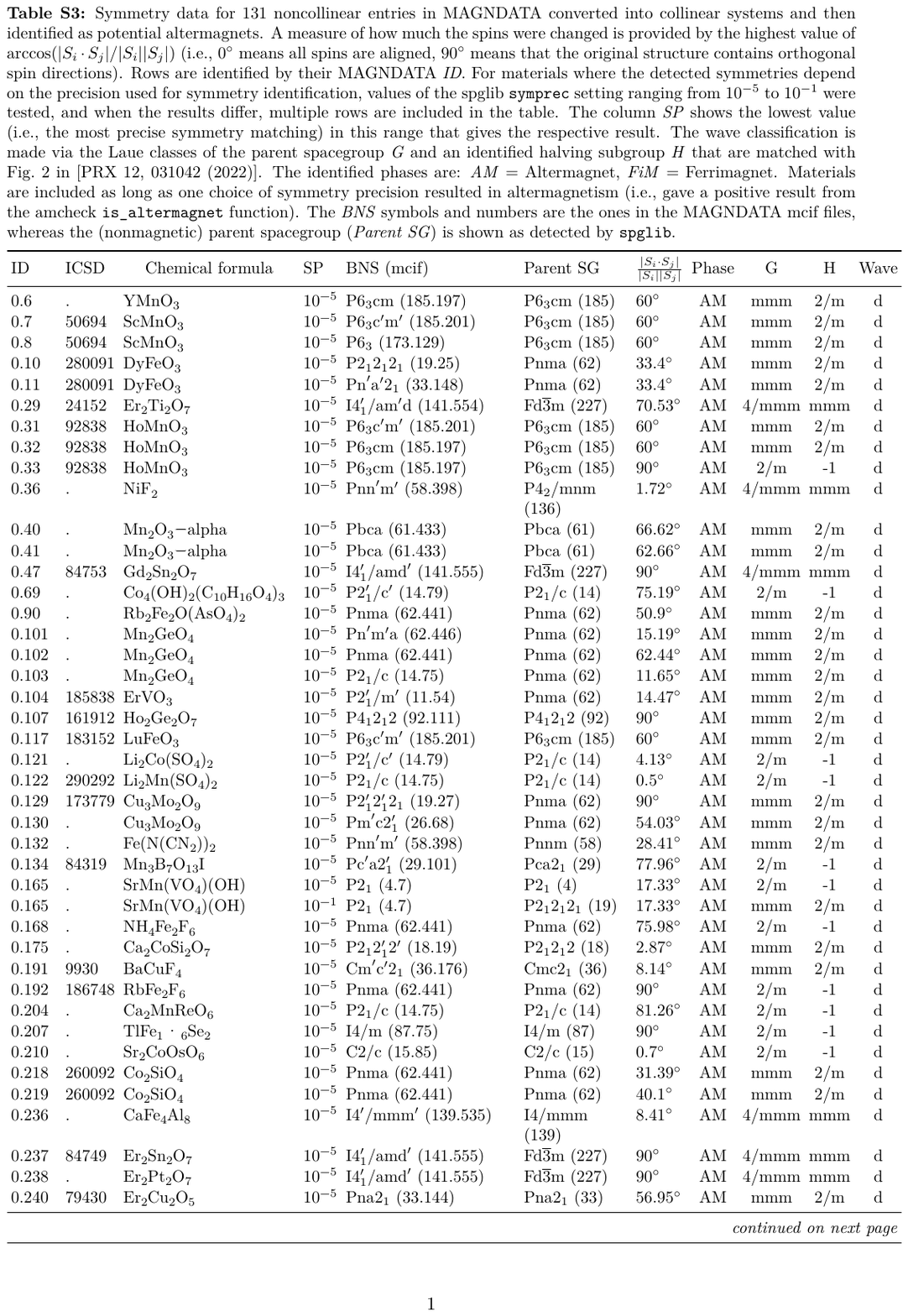}
\includepdf[pages=2]{TableS3.pdf}
\includepdf[pages=3]{TableS3.pdf}
\includepdf[pages=4]{TableS3.pdf}

\clearpage

\begin{table*}[t]
\normalsize
\caption{Sensitivity of the band gap $E_g$, maximum splitting $\Delta_{\max}$, and average splitting $\Delta_{\mathrm{avg}}$ to $\pm10\%$ variations of the Hubbard $U$ values (relative to the reference values used in the main text). Band gaps are reported with two decimal places.}
\label{tab:Uscan_pm10}
\centering
\normalsize 
\setlength{\tabcolsep}{5pt}
\renewcommand{\arraystretch}{1.08}
\begin{tabular}{l rr rr rr}
\hline\hline
& \multicolumn{2}{c}{$E_g$ (eV)} & \multicolumn{2}{c}{$\Delta_{\max}$ (meV)} & \multicolumn{2}{c}{$\Delta_{\mathrm{avg}}$ (meV)} \\
\cline{2-3}\cline{4-5}\cline{6-7}
Material & $U{-}10\%$ & $U{+}10\%$ & $U{-}10\%$ & $U{+}10\%$ & $U{-}10\%$ & $U{+}10\%$ \\
\hline
Ba$_3$CoIr$_2$O$_9$            & 0.00 & 0.00 &  258.6 &  221.0 & 172.69 & 151.41 \\
Ca$_2$MnReO$_6$                & 0.00 & 0.00 &  145.8 &  143.2 &  77.69 &  75.11 \\
CaFe$_4$Al$_8$                 & 0.00 & 0.00 &  606.8 &  612.3 & 368.73 & 382.87 \\
Ce$_4$Sb$_3$                   & 0.00 & 0.00 &  152.2 &  152.2 &  34.21 &  34.21 \\
CeMn$_2$Ge$_4$O$_{12}$         & 0.00 & 0.00 &  140.4 &  164.1 & 115.22 & 121.32 \\
Co$_2$Mo$_3$O$_8$              & 0.00 & 1.65 &   72.3 &   72.7 &  37.81 &  45.18 \\
CoTa$_4$Se$_8$                 & 0.00 & 0.00 &  158.1 &  148.8 & 126.14 & 127.04 \\
CrNb$_4$S$_8$                  & 0.00 & 0.00 &  113.2 &  111.5 &  61.36 &  55.54 \\
CrSb                           & 0.00 & 0.00 & 1797.6 & 1921.0 & 736.97 & 782.25 \\
CrSe                           & 0.00 & 0.00 &  813.7 &  787.2 & 494.50 & 485.39 \\
Cu$_2$O$_3$Cl                  & 0.00 & 0.00 &  146.6 &  157.3 & 84.62 & 104.85 \\
La$_2$CoIrO$_6$                & 0.00 & 0.00 &  269.5 &  269.7 & 124.47 & 121.21 \\
NbMnP                          & 0.00 & 0.00 &  439.8 &  478.4 & 271.28 & 278.59 \\
PrBaMn$_2$O$_6$                & 0.00 & 0.00 &  115.2 &  133.8 &  52.98 &  54.41 \\
RbCoBr$_3$                     & 0.00 & 0.00 &  144.5 &  152.6 &  94.97 &  90.57 \\
RuO$_2$                        & 0.00 & 0.00 &  865.9 &  969.4 & 344.38 & 356.02 \\
Sr$_2$CoOsO$_6$                & 0.00 & 0.00 &   67.0 &   67.6 &  41.90 &  41.59 \\
Sr$_2$ScOsO$_6$                & 0.00 & 0.00 &   94.0 &   94.0 &  45.34 &  45.34 \\
UCr$_2$Si$_2$C                 & 0.00 & 0.00 &  409.1 &  925.7 & 217.15 & 369.64 \\
UNiGa                          & 0.00 & 0.00 &   47.4 &   41.3 &  24.18 &  22.36 \\
VNb$_3$S$_6$                   & 0.00 & 0.00 &  154.1 &  169.1 &  59.18 &  60.86 \\
\hline\hline
\end{tabular}%
\end{table*}

\clearpage

\begin{figure}[]
  \centering
  \includegraphics[width=\linewidth]{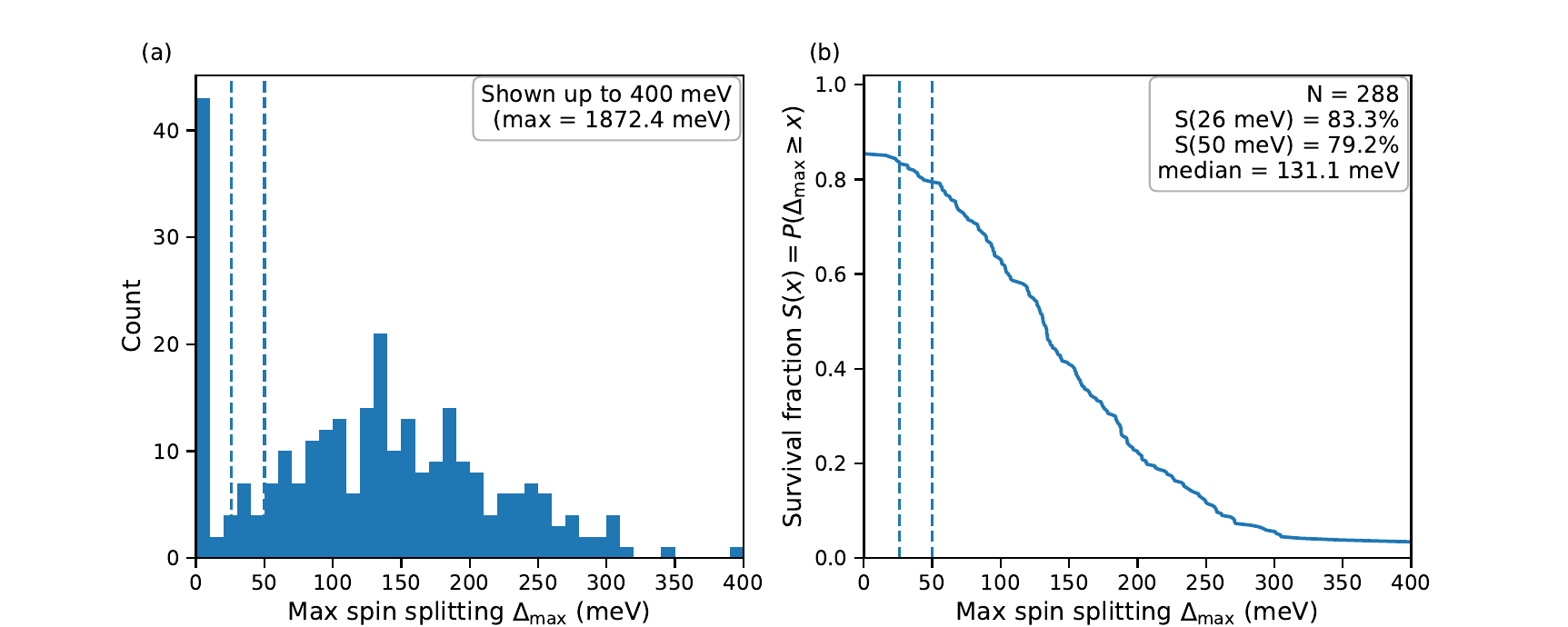}
  \caption{(a) Histogram of the maximal altermagnetic spin splitting $\Delta_{\mathrm{max}}$ across the screened set ($N=288$), shown up to 800~meV for clarity (outliers extend to 1872.4~meV). (b) Survival distribution $S(x)=P(\Delta_{\mathrm{max}}\geq x)$ with dashed lines marking 26~meV ($k_B T$ at 300~K) and the screening threshold 50~meV.}
  \label{fig:S1}
\end{figure}

\clearpage

\includepdf[pages=1]{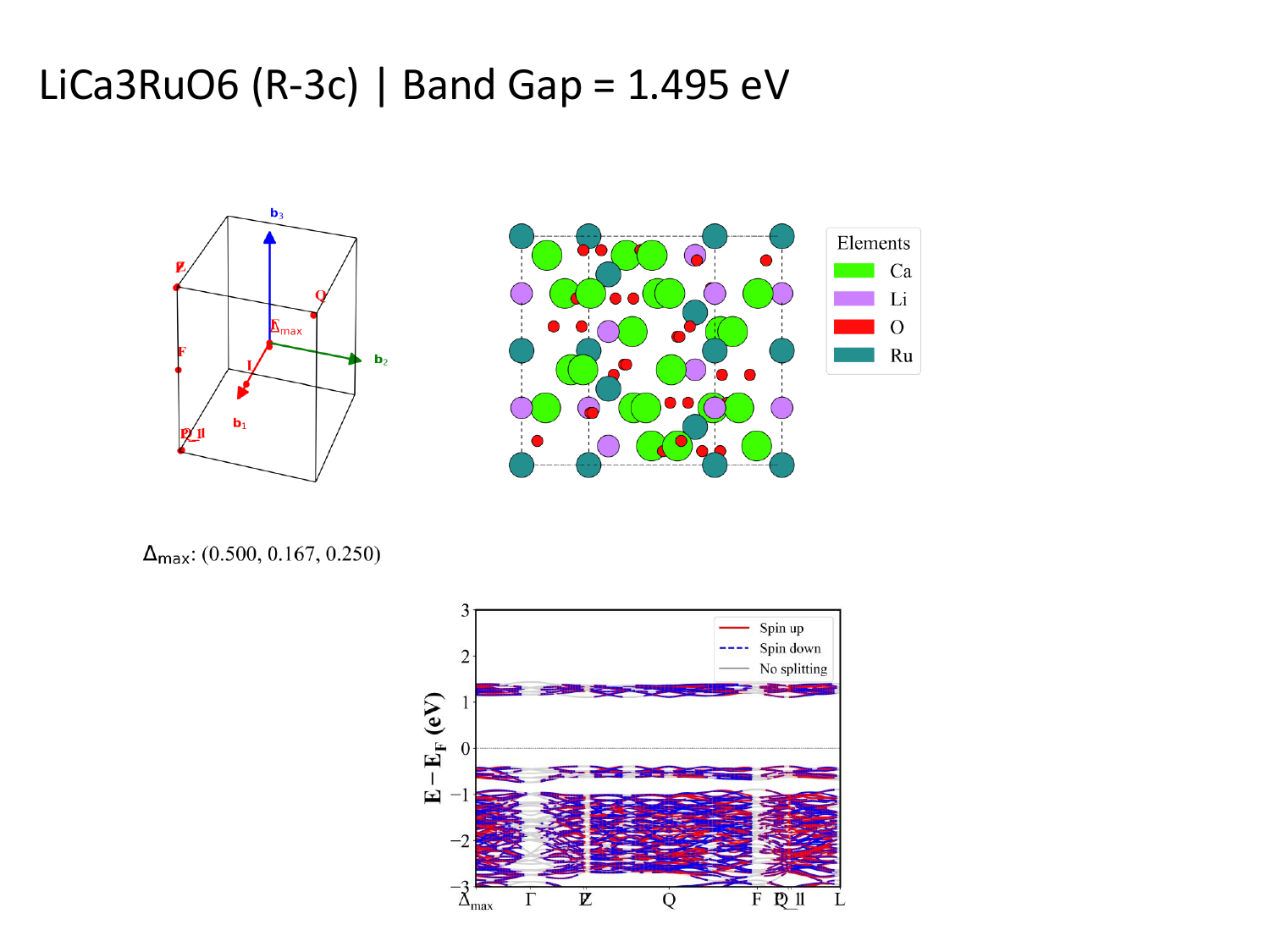}
\includepdf[pages=2]{S1.pdf}
\includepdf[pages=3]{S1.pdf}
\includepdf[pages=4]{S1.pdf}
\includepdf[pages=5]{S1.pdf}
\includepdf[pages=6]{S1.pdf}
\includepdf[pages=7]{S1.pdf}
\includepdf[pages=8]{S1.pdf}
\includepdf[pages=9]{S1.pdf}
\includepdf[pages=10]{S1.pdf}
\includepdf[pages=11]{S1.pdf}
\includepdf[pages=12]{S1.pdf}
\includepdf[pages=13]{S1.pdf}
\includepdf[pages=14]{S1.pdf}
\includepdf[pages=15]{S1.pdf}
\includepdf[pages=16]{S1.pdf}
\includepdf[pages=17]{S1.pdf}
\includepdf[pages=18]{S1.pdf}
\includepdf[pages=19]{S1.pdf}

\includepdf[pages=1]{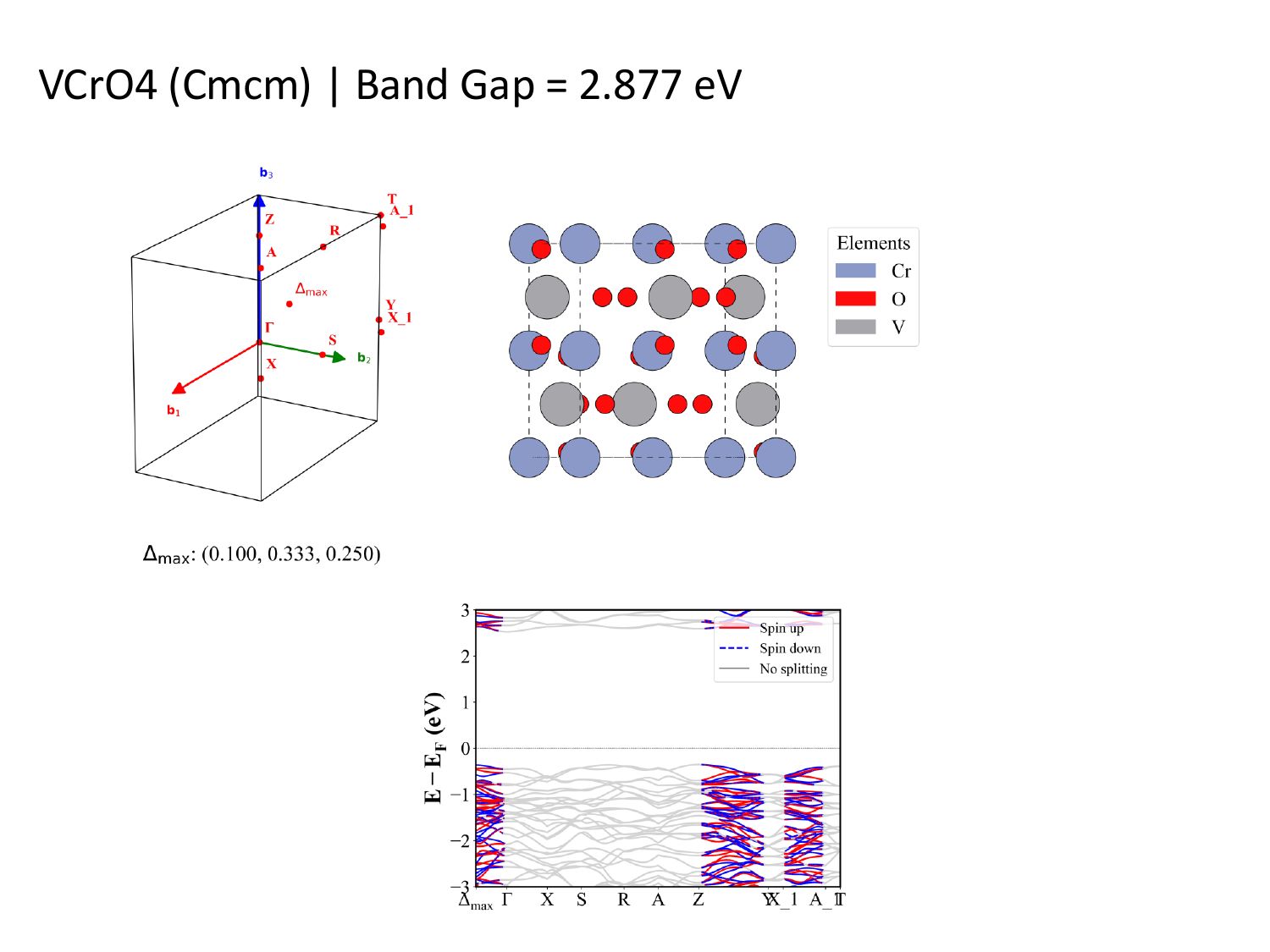}
\includepdf[pages=2]{S2.pdf}
\includepdf[pages=3]{S2.pdf}
\includepdf[pages=4]{S2.pdf}
\includepdf[pages=5]{S2.pdf}
\includepdf[pages=6]{S2.pdf}
\includepdf[pages=7]{S2.pdf}
\includepdf[pages=8]{S2.pdf}
\includepdf[pages=9]{S2.pdf}
\includepdf[pages=10]{S2.pdf}
\includepdf[pages=11]{S2.pdf}
\includepdf[pages=12]{S2.pdf}
\includepdf[pages=13]{S2.pdf}
\includepdf[pages=14]{S2.pdf}
\includepdf[pages=15]{S2.pdf}
\includepdf[pages=16]{S2.pdf}
\includepdf[pages=17]{S2.pdf}
\includepdf[pages=18]{S2.pdf}
\includepdf[pages=19]{S2.pdf}
\includepdf[pages=20]{S2.pdf}

\includepdf[pages=1]{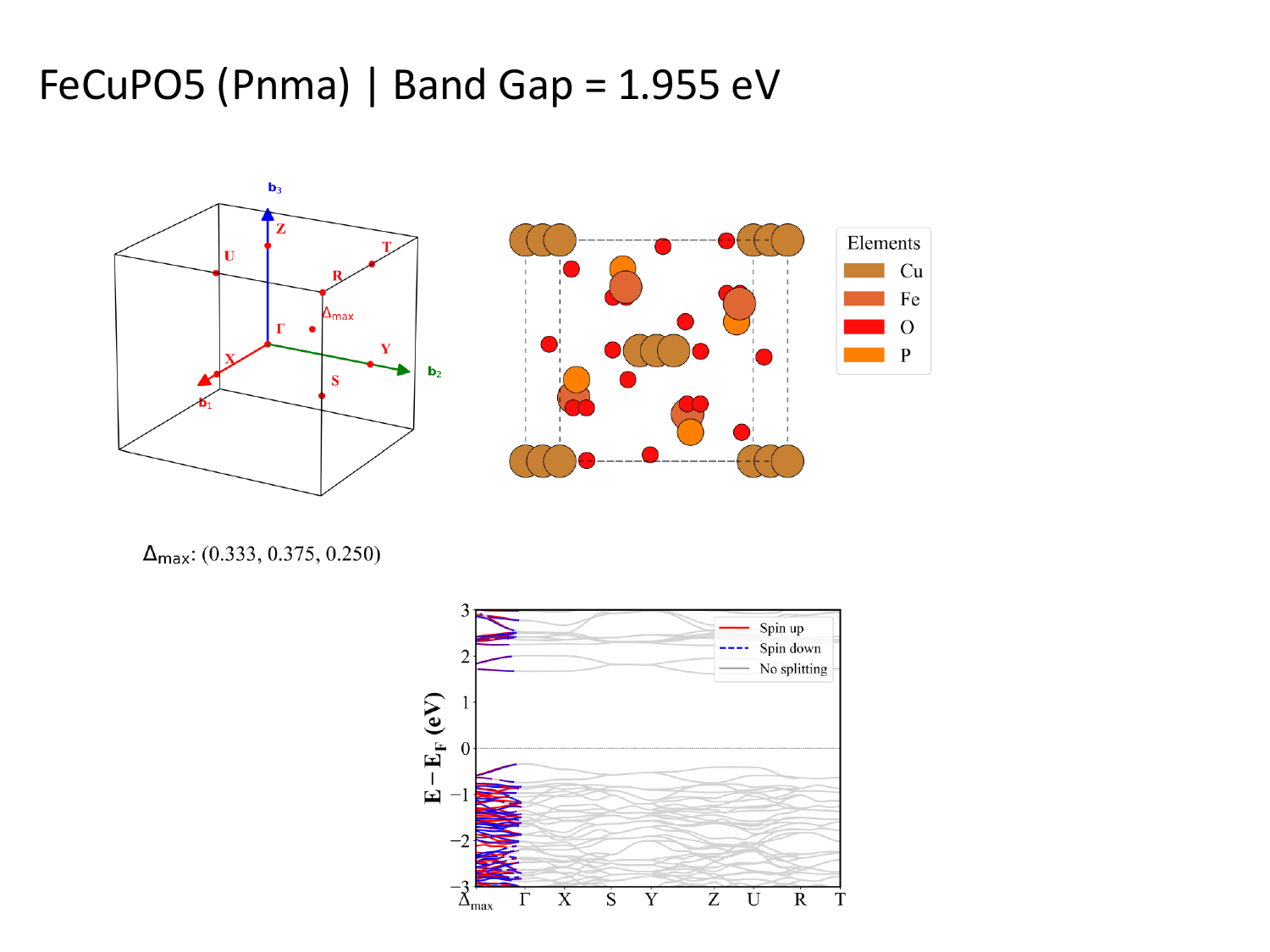}
\includepdf[pages=2]{S3.pdf}
\includepdf[pages=3]{S3.pdf}
\includepdf[pages=4]{S3.pdf}
\includepdf[pages=5]{S3.pdf}
\includepdf[pages=6]{S3.pdf}
\includepdf[pages=7]{S3.pdf}
\includepdf[pages=8]{S3.pdf}
\includepdf[pages=9]{S3.pdf}
\includepdf[pages=10]{S3.pdf}
\includepdf[pages=11]{S3.pdf}
\includepdf[pages=12]{S3.pdf}
\includepdf[pages=13]{S3.pdf}

\includepdf[pages=1]{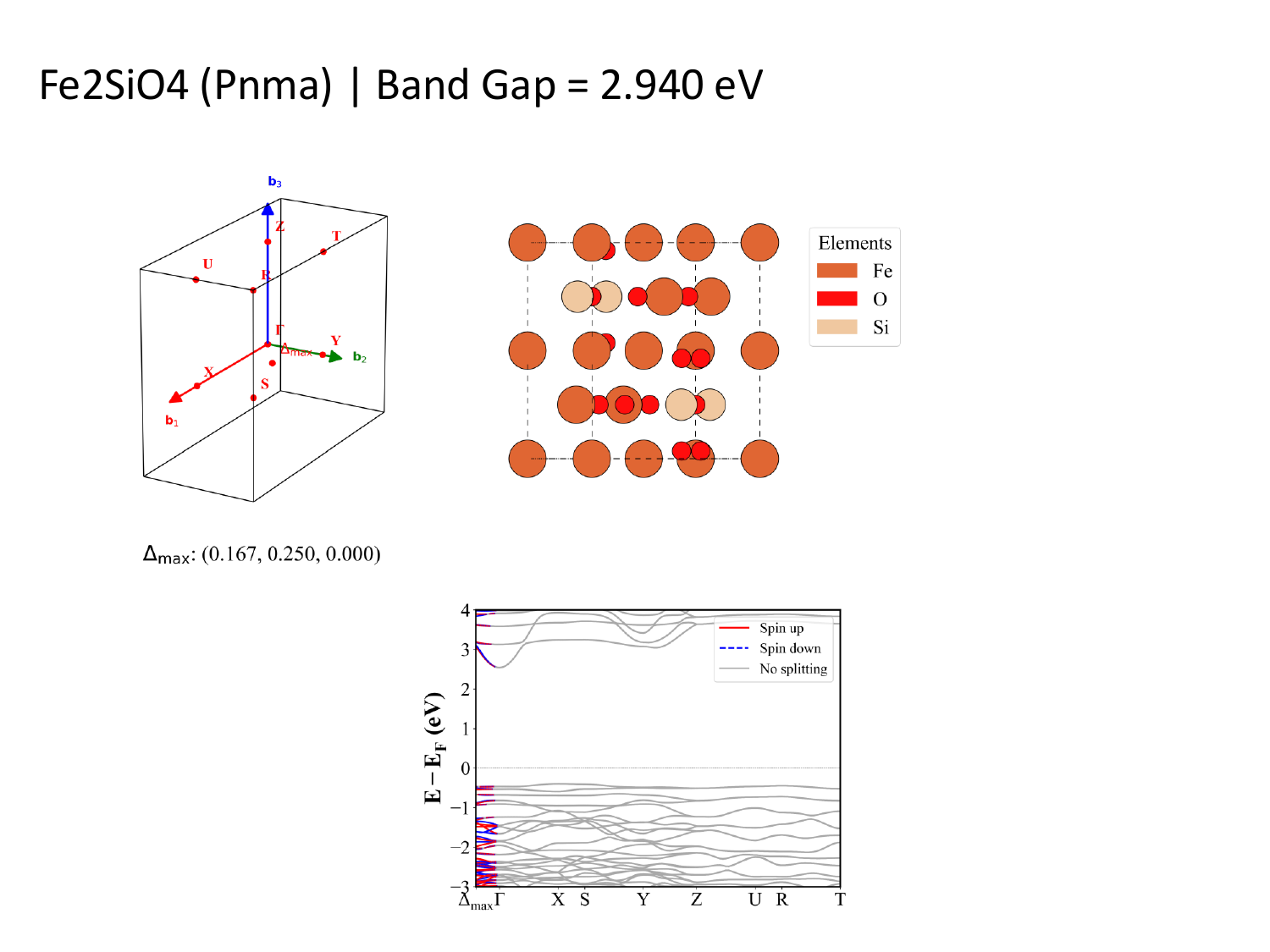}
\includepdf[pages=2]{S4.pdf}
\includepdf[pages=3]{S4.pdf}
\includepdf[pages=4]{S4.pdf}
\includepdf[pages=5]{S4.pdf}
\includepdf[pages=6]{S4.pdf}
\includepdf[pages=7]{S4.pdf}
\includepdf[pages=8]{S4.pdf}
\includepdf[pages=9]{S4.pdf}
\includepdf[pages=10]{S4.pdf}
\includepdf[pages=11]{S4.pdf}
\includepdf[pages=12]{S4.pdf}
\includepdf[pages=13]{S4.pdf}

\includepdf[pages=1]{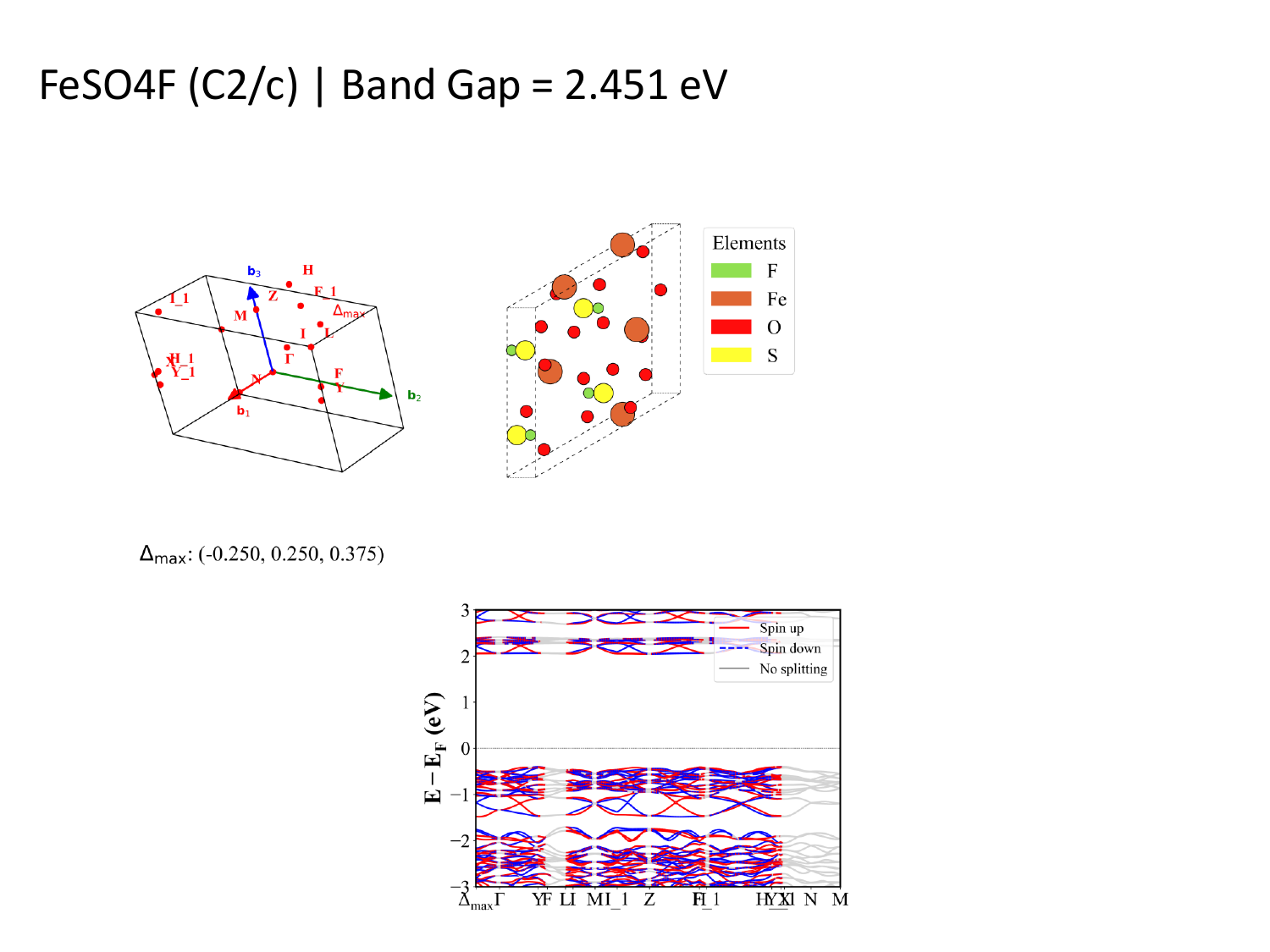}
\includepdf[pages=2]{S5.pdf}
\includepdf[pages=3]{S5.pdf}
\includepdf[pages=4]{S5.pdf}
\includepdf[pages=5]{S5.pdf}
\includepdf[pages=6]{S5.pdf}
\includepdf[pages=7]{S5.pdf}
\includepdf[pages=8]{S5.pdf}
\includepdf[pages=9]{S5.pdf}
\includepdf[pages=10]{S5.pdf}
\includepdf[pages=11]{S5.pdf}
\includepdf[pages=12]{S5.pdf}
\includepdf[pages=13]{S5.pdf}
\includepdf[pages=14]{S5.pdf}
\includepdf[pages=15]{S5.pdf}

\includepdf[pages=1]{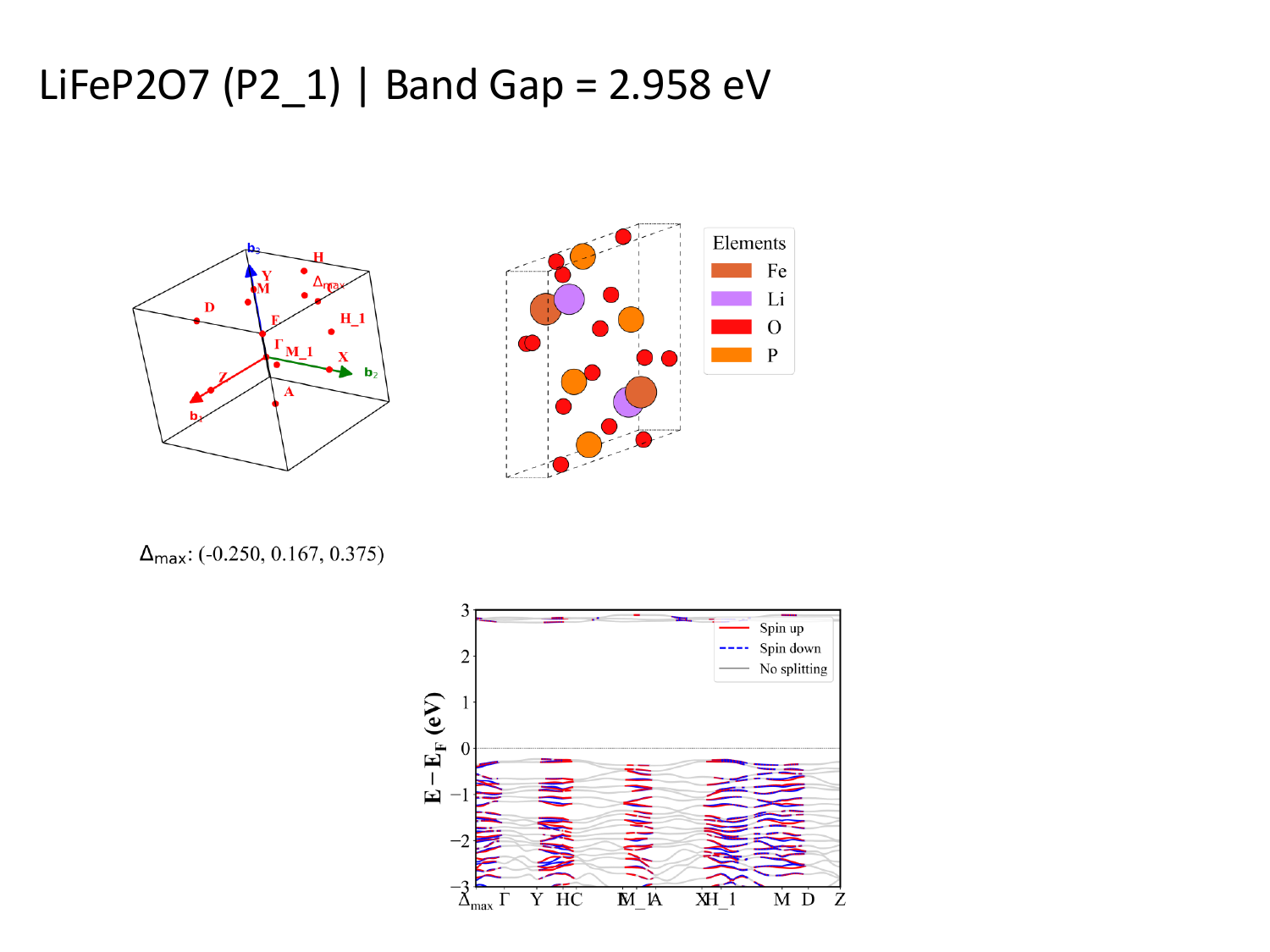}
\includepdf[pages=2]{S6.pdf}
\includepdf[pages=3]{S6.pdf}
\includepdf[pages=4]{S6.pdf}
\includepdf[pages=5]{S6.pdf}
\includepdf[pages=6]{S6.pdf}
\includepdf[pages=7]{S6.pdf}
\includepdf[pages=8]{S6.pdf}
\includepdf[pages=9]{S6.pdf}
\includepdf[pages=10]{S6.pdf}
\includepdf[pages=11]{S6.pdf}
\includepdf[pages=12]{S6.pdf}
\includepdf[pages=13]{S6.pdf}
\includepdf[pages=14]{S6.pdf}
\includepdf[pages=15]{S6.pdf}
\includepdf[pages=16]{S6.pdf}
\includepdf[pages=17]{S6.pdf}
\includepdf[pages=18]{S6.pdf}
\includepdf[pages=19]{S6.pdf}
\includepdf[pages=20]{S6.pdf}
\includepdf[pages=21]{S6.pdf}
\includepdf[pages=22]{S6.pdf}

\includepdf[pages=1]{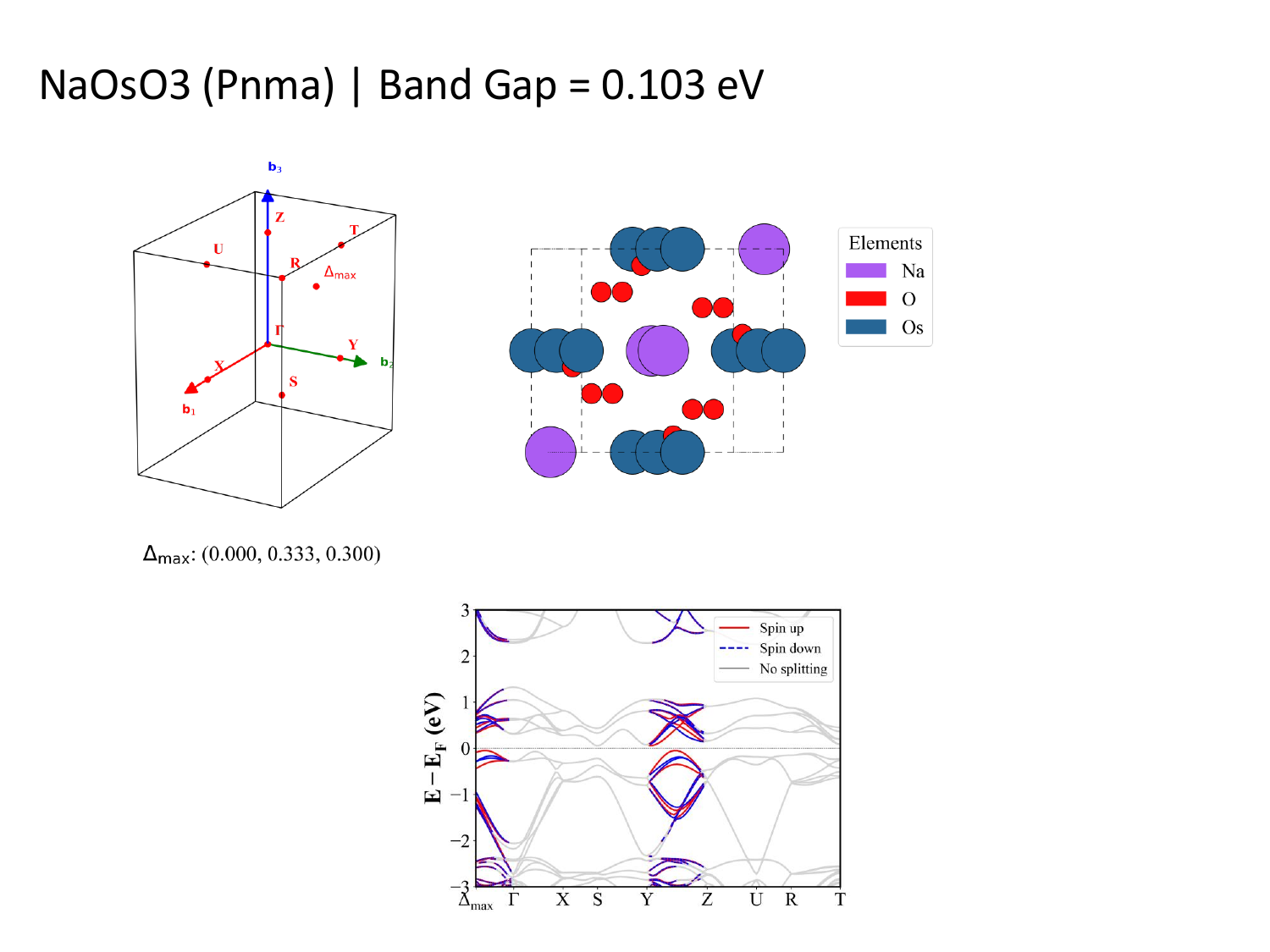}
\includepdf[pages=2]{S7.pdf}
\includepdf[pages=3]{S7.pdf}
\includepdf[pages=4]{S7.pdf}
\includepdf[pages=5]{S7.pdf}
\includepdf[pages=6]{S7.pdf}
\includepdf[pages=7]{S7.pdf}
\includepdf[pages=8]{S7.pdf}
\includepdf[pages=9]{S7.pdf}
\includepdf[pages=10]{S7.pdf}
\includepdf[pages=11]{S7.pdf}
\includepdf[pages=12]{S7.pdf}
\includepdf[pages=13]{S7.pdf}
\includepdf[pages=14]{S7.pdf}
\includepdf[pages=15]{S7.pdf}
\includepdf[pages=16]{S7.pdf}
\includepdf[pages=17]{S7.pdf}
\includepdf[pages=18]{S7.pdf}

\includepdf[pages=1]{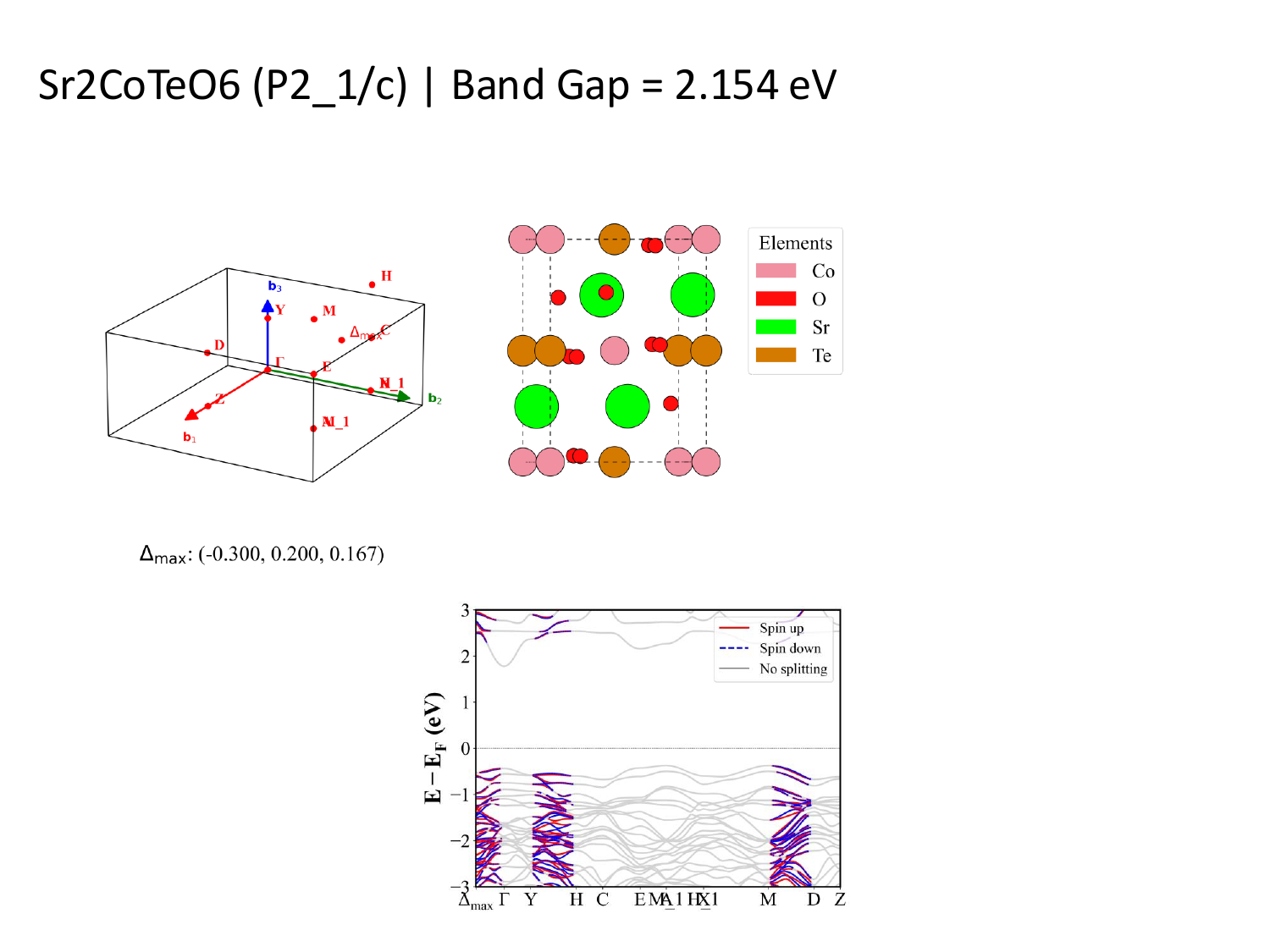}
\includepdf[pages=2]{S8.pdf}
\includepdf[pages=3]{S8.pdf}
\includepdf[pages=4]{S8.pdf}
\includepdf[pages=5]{S8.pdf}
\includepdf[pages=6]{S8.pdf}
\includepdf[pages=7]{S8.pdf}
\includepdf[pages=8]{S8.pdf}
\includepdf[pages=9]{S8.pdf}
\includepdf[pages=10]{S8.pdf}
\includepdf[pages=11]{S8.pdf}
\includepdf[pages=12]{S8.pdf}
\includepdf[pages=13]{S8.pdf}
\includepdf[pages=14]{S8.pdf}
\includepdf[pages=15]{S8.pdf}
\includepdf[pages=16]{S8.pdf}
\includepdf[pages=17]{S8.pdf}
\includepdf[pages=18]{S8.pdf}
\includepdf[pages=19]{S8.pdf}
\includepdf[pages=20]{S8.pdf}

\includepdf[pages=1]{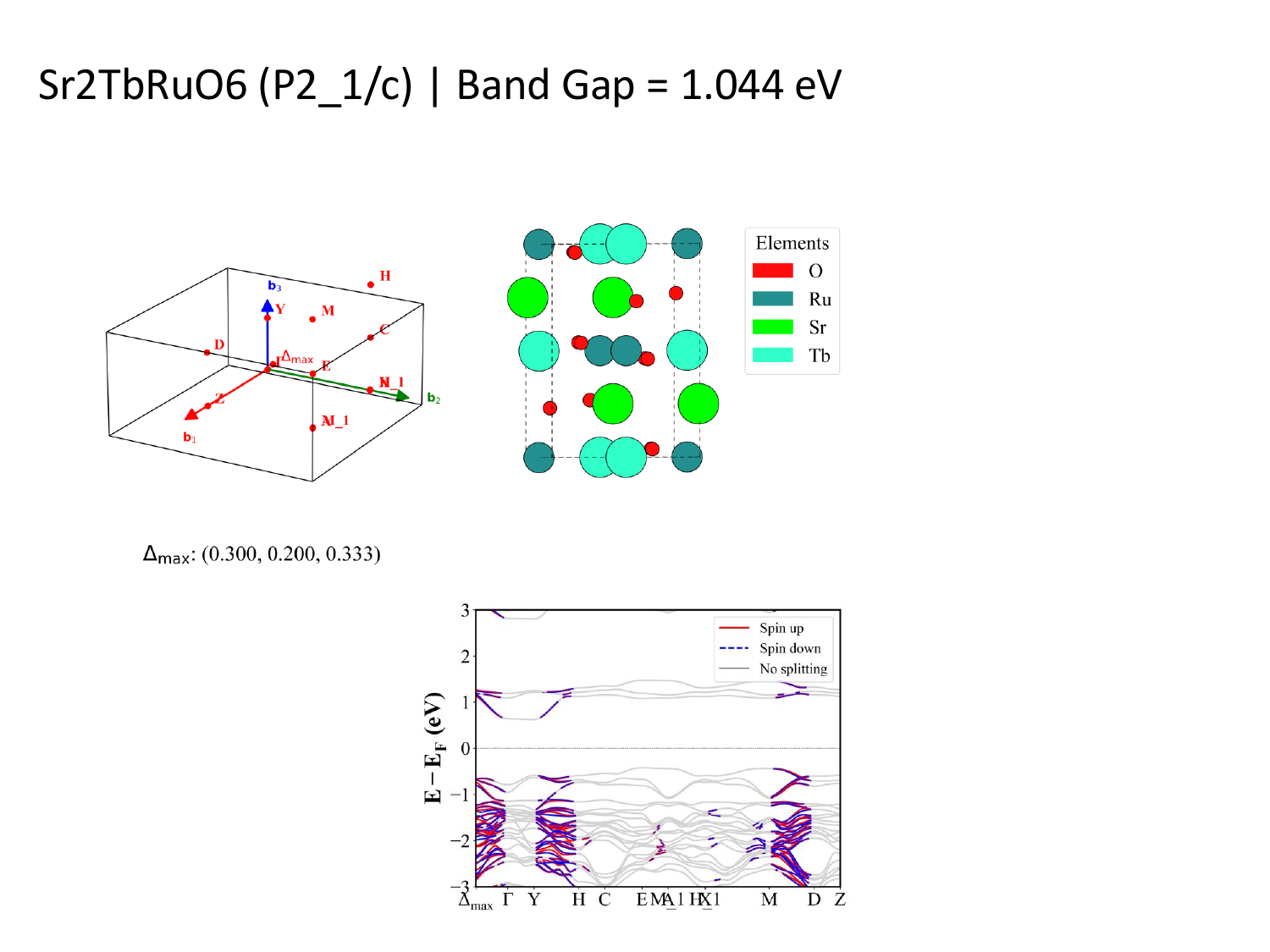}
\includepdf[pages=2]{S9.pdf}
\includepdf[pages=3]{S9.pdf}
\includepdf[pages=4]{S9.pdf}
\includepdf[pages=5]{S9.pdf}
\includepdf[pages=6]{S9.pdf}
\includepdf[pages=7]{S9.pdf}
\includepdf[pages=8]{S9.pdf}
\includepdf[pages=9]{S9.pdf}
\includepdf[pages=10]{S9.pdf}
\includepdf[pages=11]{S9.pdf}
\includepdf[pages=12]{S9.pdf}
\includepdf[pages=13]{S9.pdf}
\includepdf[pages=14]{S9.pdf}
\includepdf[pages=15]{S9.pdf}
\includepdf[pages=16]{S9.pdf}
\includepdf[pages=17]{S9.pdf}

\includepdf[pages=1]{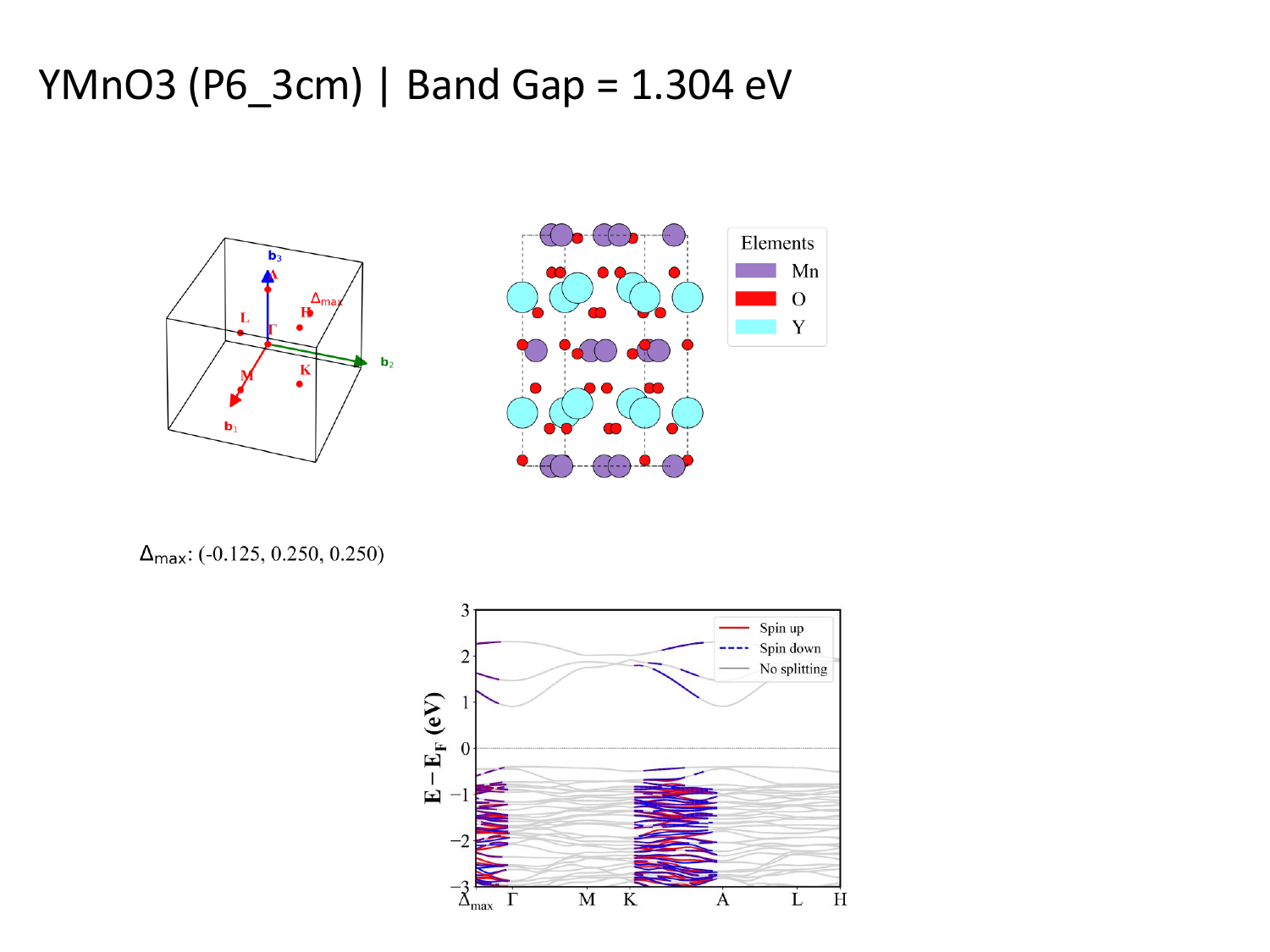}
\includepdf[pages=2]{S10.pdf}
\includepdf[pages=3]{S10.pdf}
\includepdf[pages=4]{S10.pdf}
\includepdf[pages=5]{S10.pdf}
\includepdf[pages=6]{S10.pdf}
\includepdf[pages=7]{S10.pdf}
\includepdf[pages=8]{S10.pdf}
\includepdf[pages=9]{S10.pdf}
\includepdf[pages=10]{S10.pdf}
\includepdf[pages=11]{S10.pdf}
\includepdf[pages=12]{S10.pdf}
\includepdf[pages=13]{S10.pdf}
\includepdf[pages=14]{S10.pdf}
\includepdf[pages=1]{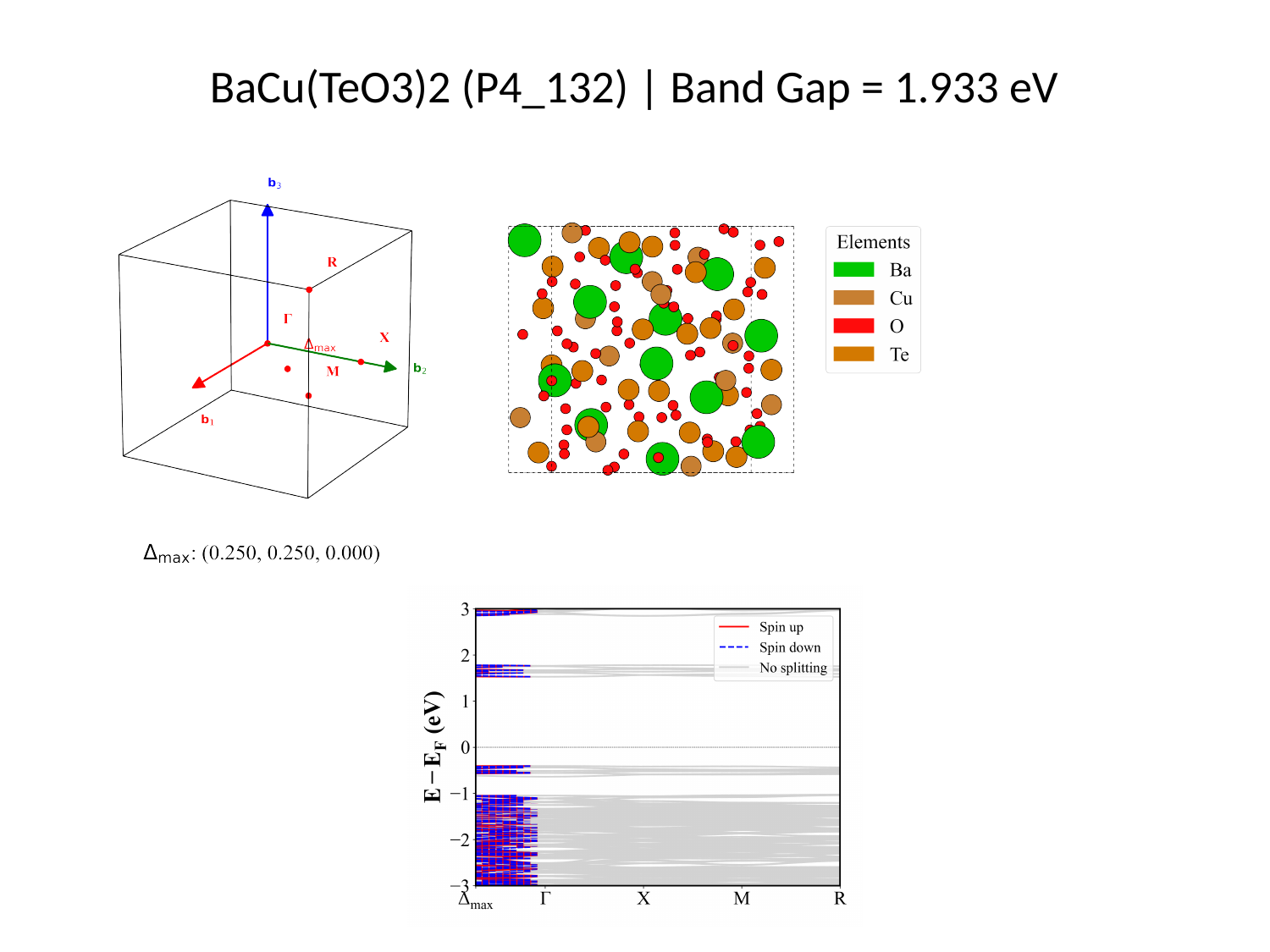}
\includepdf[pages=2]{S11.pdf}
\includepdf[pages=3]{S11.pdf}
\includepdf[pages=4]{S11.pdf}
\includepdf[pages=5]{S11.pdf}
\includepdf[pages=6]{S11.pdf}
\includepdf[pages=7]{S11.pdf}
\includepdf[pages=8]{S11.pdf}
\includepdf[pages=9]{S11.pdf}
\includepdf[pages=10]{S11.pdf}

\end{document}